# Canonically Relativistic Quantum Mechanics: Casimir Field Equations of the Quaplectic Group

*Stephen G. Low*[1]

The Hilbert space of the unitary irreducible representations of a Lie group that is a quantum dynamical symmetry group are identified with the physical quantum state space. Hermitian representation of the algebra are observables. The eigenvalue equations for the representation of the set of Casimir invariant operators define the field equations of the system.

A general class of dynamical groups are semidirect products $\mathcal{K} \otimes_s \mathcal{N}$ for which the representations are given by Mackey's theory. The homogeneous group $\mathcal{K}$ must be a subgroup of the automorphisms of the normal group $\mathcal{N}$.

The archetype dynamical symmetry group is the Poincaré group. The hyperbolic rotations of Lorentz homogeneous group on position, time or momentum, energy spaces embody special relativity The field equations defined by the representations of the Casimir operators define the basic equations of physics; Klein-Gordon, Dirac, Maxwell and so forth.

This paper explores a more general dynamical group candidate that is also a semi-direct product but where the *translation* normal subgroup $\mathcal{N}$ is now the Heisenberg group (which itself is the semidirect product of translation groups). The relevant automorphisms of the Heisenberg group are the symplectic group. This together with the requirement for an orthogonal metric leads to the pseudo-unitary group for the homogeneous group $\mathcal{K}$. This group acts on a nonabelian position, time, energy, momentum space. It contains four Poincaré subgroups, two associated with special (velocity) relativity and two defining a similar relativity principle that generalizes the concept of force to hyperbolic rotations on the momentum, time and energy, position subspaces. As it is derived from the symplectic group, contains four (quad) Poincaré groups and has the quantum *translations*, we call it the *quaplectic* group[2]. The theory embodies Born's reciprocity and a new relativity principle.

The physical meaning and motivation of the quaplectic group is presented and the Hermitian irreducible representations of the algebra are, as in the Poincaré case, determined using Mackey's theory of representations of semidirect products. As with the Poincaré group, choice of the group defines the Hilbert space of representations that are identified with quantum particle states. The field equations, the representation of the Casimir operators, are obtained and investigated.

# 1: Introduction

## 1.1 Dynamical groups

The unitary irreducible representations $\varrho$ of a Lie group $\mathcal{G}$ act as unitary operators on a Hilbert space $\mathsf{H}^\varrho$

$$\varrho(g): \mathsf{H}^\varrho \to \mathsf{H}^\varrho : |\psi\rangle \mapsto \varrho(g)|\psi\rangle \text{ where } g \in \mathcal{G} \text{ and } \varrho(g)^\dagger = \varrho(g)^{-1}.$$

The Hilbert space $\mathsf{H}^\varrho$ is determined by the group and the unitary irreducible representation $\varrho$. The Lie algebra of the group may be identified with the tangent space of the group $\mathfrak{a}(\mathcal{G}) \simeq T_e \mathcal{G}$ and the lift $\varrho'$ of the representation to the algebra represents elements of the algebra $X \in \mathfrak{a}(\mathcal{G})$ as anti-Hermitian operators on $\mathsf{H}^\varrho$

$$\varrho'(X): \mathsf{H}^\varrho \to \mathsf{H}^\varrho : |\psi\rangle \mapsto \varrho'(X)|\psi\rangle \text{ with } X \in \mathfrak{a}(\mathcal{G}) \text{ and } \varrho'(X)^\dagger = -\varrho'(X).$$

Quantum mechanics defines particles states $|\psi\rangle$ to be elements of a Hilbert space $\mathsf{H}$. Observables are Hermitian operators $H = H^\dagger$, $H: \mathsf{H} \to \mathsf{H} : |\psi\rangle \mapsto H|\psi\rangle$ and unitary operators $U^\dagger = U^{-1}$ defines the evolution of the states. A complete orthonormal basis $|\varphi_x\rangle$ satisfies $\langle \varphi_{\tilde{x}} | \varphi_x \rangle = \delta_{\tilde{x},x}$ and $|\varphi_x\rangle\langle\varphi_{\tilde{x}}| = \mathbb{1}$ where $\mathbb{1}$ is the identity on $\mathsf{H}$. Observable matrix elements are $h_{\tilde{x},x} = \langle \varphi_{\tilde{x}} | H | \varphi_x \rangle$ with $h_{\tilde{x},x} \in \mathbb{R}$. A unitary operator $U$ transforms orthonormal bases into orthonormal bases $|\tilde{\varphi}_x\rangle = U|\varphi_x\rangle$ and hence preserves probabilities $\langle \tilde{\varphi}_{\tilde{x}} | \tilde{\varphi}_x \rangle = \langle \tilde{\varphi}_{\tilde{x}} | U^\dagger U | \varphi_x \rangle = \langle \tilde{\varphi}_{\tilde{x}} | \varphi_x \rangle$.

We will say that $\mathcal{G}$ is a dynamical symmetry of a quantum system if the following conditions are met. First, the Hilbert space $\mathsf{H}^\varrho$ determined by the unitary irreducible representations $\varrho$ of a group $\mathcal{G}$ is identical to the Hilbert space $\mathsf{H}$ of the quantum system in question, $\mathsf{H} \simeq \mathsf{H}^\varrho$. This means that the (particle) states of the quantum system are unitary irreducible representations of the dynamical group. Unitary operators $U$ defining unitary evolution of the system are given by $U = \varrho(g)$ with corresponding observables $\mathsf{H} = i\, \varrho'(X)$. Note that the $i$ is required to provide the one to one mapping of the anti-Hermitian operators that represent the elements of the Lie algebra with the Hermitian observables of standard quantum mechanics[3]. Key physical invariant quantities are given by the Casimir invariant operators and the eigenvalues label the unitary irreducible representations.

Casimir invariant operators $C_i$ are elements of the enveloping algebra of the Lie algebra of the group, $\mathfrak{e}(\mathcal{G})$, that commute with all elements of the algebra; $[C_\alpha, X] = 0$ for all $X \in \mathfrak{a}(\mathcal{G})$. The number of Casimir invariants is $\alpha = 1, 2, \ldots N_c$ where $N_c = N_g - N_r$ is the number of Casimir invariants with $N_g$ is the dimension of the group and $N_r$ is its rank[4]. The unitary irreducible representation $\varrho'(C_\alpha)$ have eigenvalues $c_\alpha$ that are constants for a given unitary irreducible representation.

$$\varrho'(C_\alpha)|\psi\rangle = c_\alpha|\psi\rangle \text{ with } |\psi\rangle \in \mathsf{H}^\varrho \ , \alpha = 1, 2...N_c$$

As this is the representation of the Casimir invariant $C_\alpha$ as a Hermitian operator $\varrho'(C_\alpha)$ on the Hilbert space, the eigenvalues $c_\alpha$ are real. For certain groups such as the compact semi-simple groups, the eigenvalues actually label the unitary irreducible representations uniquely. This is not generally true but it is always true that the above eigenvalue equation hold for a given unitary irreducible representation. These equations in the physical theory are the *field equations* for the dynamical group. The simultaneous solution of these eigenvalue equations define the particle states of the theory and the eigenvalues define physically observable, constant properties that are attributed to these particle states.

The group, representations and field equations must contract in a limit to effectively the dynamical group characterizing quantum systems in a known limit. The free particle, relativistic case is the Poincaré group (which in turn contracts to the Galilei group in the low velocity limit $c \to \infty$.

Our program is now straightforward. Choose a dynamical Lie group based on physical considerations. Determine the unitary irreducible representations $\varrho$ of the group and the Hilbert space $\mathsf{H}^\varrho$ and the eigenvalue equations of the unitary irreducible representations of the Casimir invariants. This is a strictly mathematical exercise that follows deterministically from the choice of the dynamical Lie group. Determine whether the resulting field equations (the representations of the Casimir invariants eigenvalue equations) and Hilbert space of observables corresponds to experimentally observed phenomena.

We must immediately note that the problem of determining the unitary irreducible representations of an arbitrary Lie group, while in principle solvable, is not a generally solved problem in the mathematical literature. Fortunately, the class of groups that we are interested in has been explicitly solved, although the calculations can be extraordinarily complex as the dimensions of the groups in question increase. This class of groups of interest is the direct and semi-direct products of the closed subgroups of the real general linear Lie Groups. Furthermore, all groups of interest are algebraic[5] in the sense that they are defined by simple polynomial constraints on the general linear group. This includes all of the compact and non-compact semi-simple real groups, the translation group, the Heisenberg group and various direct and semi-direct products of these groups. The unitary irreducible representations of the translation group are straightforward and the unitary irreducible representations of the *classical* semi-simple groups of interest, both compact and noncompact, have been essentially determined. Computation of the direct product representations are trivial. The condition that the groups matrix groups that are algebraic is a sufficient condition to enable the Mackey theory [29,28,32,49] of semi-direct product representations to be applied to construct the remainder of the unitary irreducible representations in this class of groups including the Heisenberg groups (that is the semidirect product of two translation groups) and semidirect products constructed from various combinations of the translation, Heisenberg groups and other semisimple groups.

## 1.2 The Poincaré dynamical group

A dynamical group in the above class may be chosen and the mathematical machinery gives the structure of the representation Hilbert space, that is identified with the quantum particle state space, and the eigenvalue equation for the representations of the Casimir operators that define the particle state field equations.

The remaining problem is to determine the dynamical group of physical interest.

The archetype for a dynamical group, as defined above, is clearly the Poincaré group[6]. The eigenvalues of the unitary irreducible representations of the Casimir invariants of this group define the physical concepts of mass and spin (or helicity) and the simultaneous solution of the corresponding eigenvalue equations correspond to the field equations that define many of the celebrated equations in physics, Klein-Gordon, Dirac, Maxwell, and so forth.

So the question is: are we done. Is there any reason to look at other dynamical groups. This is the central question that is examined in this paper.

As noted, the unitary irreducible representations of the Poincaré group, (i.e. the universal cover[7] of $\mathcal{E}(1, n) = \mathcal{SO}(1, n) \otimes_s \mathcal{T}(1, n)$ ), where $\mathcal{SO}(1, n)$ is the special pseudo-orthogonal group and $\mathcal{T}(1, n) \simeq \mathcal{T}(1 + n)$[8] is the translation group, are remarkable successful in characterizing free particle states. There are however, problems which suggest that the theory may be a limiting approximation of a more general dynamical group.

The first problem is associated with the fact that the theory predicts a continuous mass spectrum that we do not observe in nature. Closely associated is that the mass is a Casimir invariant eigenvalue that is constant on a particle state (that is, an irreducible representation) and there is no mechanism in the theory for there to be transitions between irreducible states.

This shows up in the Mackey representation theory of the Poincaré group in that the Hilbert spaces $\mathcal{H}^\rho$ of this theory are the $\mathcal{L}^2$ spaces over the null $\mathbb{H}_0^n$, time-like $\mathbb{H}_-^n$ and spacelike $\mathbb{H}_+^n$ hyperboloids, $\mathcal{H}^\rho \simeq \mathfrak{V}^N \otimes \mathcal{L}^2(\mathbb{H}_{\pm,0}^n, \mathbb{R})$. The hyperboloids are the quotient of the homogeneous group $\mathcal{SO}(1, n)$ with the respective *little* groups $\mathcal{SO}(n)$, $\mathcal{SO}(1, n-1)$ and $\mathcal{E}(1, n-2)$. $\mathfrak{V}^N$ is the finite, (or countably infinite in the spacelike case) dimensional vector space for the unitary irreducible representations of the corresponding little groups.

$$\mathbb{H}_-^n \simeq \mathcal{SO}(1, n)/\mathcal{SO}(n),$$
$$\mathbb{H}_+^n \simeq \mathcal{SO}(1, n)/\mathcal{SO}(1, n-1),$$
$$\mathbb{H}_0^n \simeq \mathcal{SO}(1, n)/\mathcal{E}(1, n-2)$$

The relation of these hyperboloids to Minkowski space is as follows. Minkowski space may be identified with the symmetric space $\mathbb{M}^n = \mathcal{E}(1, n)/\mathcal{SO}(1, n) \simeq \mathbb{R}^n$. The Hilbert spaces are over the hyperboloids $\mathbb{H}_{\pm,0}^n$ that may be regarded to be surfaces in $\mathbb{M}^n$. The first Casimir invariant $c_2$ (that is the mass squared in the timelike cases $\mathbb{H}_-^n$) labels the distinct hyperboloids. If one takes the union of all the $\mathbb{H}_{\pm,0}^n$ over the values of $c_2$, the hyperboloids exactly fill $\mathbb{M}^n$. That is, each point in $\mathbb{M}^n$ is on one and only one $\mathbb{H}_{\pm,0}^n$ for some value of $c_2$.

Physically, this means that a massive particle is confined to the *infinitesimally thin* time-like mass-shell and there are no transitions in the theory that allow a decay or interaction changing the mass. Furthermore, the mass labeling these hyperboloids are continuous and we do not observe a continuous mass spectra.

What we see here is somewhat reminiscent of a classical oscillator as apposed to a quantum oscillator. The Hilbert space for a classical oscillator is the over the energy shell which is a $n$-sphere. The energy can have any positive real value. However, for a given unitary irreducible representations, it has a fixed value. No transitions between shells of different energy are allowed. The Hilbert space for a quantum oscillator encompasses all energy states and the field equations (that is the representations of the Casimir operators for the dynamical group in question) defines discrete probabilistic states with transitions allowed. We will return to this in more detail shortly.

Another very perplexing attribute of the unitary irreducible representations of the Poincaré group is that the Heisenberg group plays no role. The application of the standard unitary representation theory does not require any mention of the Heisenberg group. Rather, one must *second quantize* the theory, almost as an afterthought. Actually, this is very closely related to the above analogy with the quantum vs classical oscillator. The dynamical group for a (non-relativistic) quantum oscillator is $C(n) = \mathcal{U}(n) \otimes_s \mathcal{H}(n)$ where $\mathcal{U}(n)$ is the $n$ dimensional unitary group and $\mathcal{H}(n)$ is the $n$ dimensional Heisenberg group, as compared to the dynamical group for a classical oscillator $\mathcal{U}(n) \otimes_s \mathcal{T}(2n)$.

The Poincaré group is also in, a sense, too small. The internal symmetries observed in nature are larger than simply the spin and helicity associated with the Poincaré little groups. This has lead to a plethora of higher dimensional theories.

Another deep concern with the unitary representation of the Poincaré group framework is the appearance of the infinities. We conjecture that these may be related to the above two problems. That is, the *infinitesimally thin* surface that defines the mass and null shells may not be physically realistic and mathematically, limiting structures of this type may cause these mathematical artifacts. The mathematics may be telling us we have taken a non-physical limit.

The most direct and popular approach to the infinities is to postulate that there is finite minimum length scale or maximum acceleration [8,9,45,44,14,31]. In fact, this is the original origin of the prescription to do the cut-offs. There is another approach that introduces an *orthogonal* relativity concept in the non-commuting quantum phase space which is quite remarkable in its mathematical properties. To establish further context to understand this, we now turn to a brief discussion of basic non-relativistic quantum mechanics.

## 1.3 The Heisenberg dynamical group

A basic dynamical symmetry is the action of the Heisenberg group $\mathcal{H}(n)$ in non-relativistic quantum mechanics. The Heisenberg group is the semi-direct product of the translation groups $\mathcal{H}(n) = \mathcal{T}(n) \otimes_s \mathcal{T}(n+1)$ and is nonabelian $[Q_i, P_i] = I$ [9] with $\{Q_i, P_i, I\} \in \mathfrak{a}(\mathcal{H}(n))$. $i, j = 1, 2...n$ and we use natural units in which $\hbar = 1$. The single Casimir invariant is $I$.

The Mackey theory may be applied to the Heisenberg group to obtain the unitary irreducible representations and from this and it follows that the Hilbert space or the representations is $\mathsf{H}^\xi \simeq \mathcal{L}^2(\mathbb{R}^n, \mathbb{C})$. The representations of the algebra $\xi'(Q_i)$ correspond to the physically observable concept of position and $\xi'(P_i)$ momentum. Both the $Q_i$ and $P_i$ are translation generators that cannot be simultaneously diagonal in a representation $\xi$ of $\mathcal{H}(n)$ on the Hilbert space $\mathsf{H}^\xi$. Note that in a Hermitian representation $\xi'$ of the algebra[3,9],

$$[\xi'(Q_i), \xi'(P_i)] = i\, \xi'(I)$$

This yields the very familiar result for the *quantum* representations in which $\xi'(I) \neq 0$

$$\langle \delta_q | \xi'(Q_i) | \psi \rangle = q_i \psi(q), \quad \langle \delta_q | \xi'(P_i) | \psi \rangle = -i \frac{\partial}{\partial q_i} \psi(q)$$

Note that if $\xi'(I) = 0$, the representation degenerates to that of $\mathcal{T}(2n)$. This is reviewed further in section 2.4. An equally valid choice of representation from the equivalence class of the representations diagonalizes the momentum generators

$$\langle \delta_p | \xi'(P_i) | \psi \rangle = p_i \psi(p), \quad \langle \delta_p | \xi'(Q_i) | \psi \rangle = i \frac{\partial}{\partial p_i} \psi(p)$$

The single Casimir invariant is $I$ and the *field equations* for this dynamical group are the single equation

$$\xi'(I) | \psi \rangle = c_1 | \psi \rangle, \quad \kappa_1 \in \mathbb{R} \neq 0$$

It is clear that both the $\xi'(Q_i)$ and $\xi'(P_i)$ generate *translations* but, of course, these translations cannot be represented by simultaneously diagonal Hermitian operators on the Hilbert space $\mathsf{H}^\xi$. However, never-the-less, in this sense, the $\mathcal{H}(n)$ group plays the role of a *translation* group, in a generalized sense, on this non abelian space.

It is clear that the Heisenberg group fulfills the requirements for the definition of a dynamical group acting on the non-relativistic state space. The Hilbert space is what one expects for basic non-relativistic quantum mechanics and the representation of the generators yields the fundamental observables of position and momentum. The Hilbert space is over all of $\mathbb{R}^n$ and not constrained to some surface within it.

The content of the field equations, is clearly lacking. The only Casimir operator is trivially $I$. One certainly does not have the rich set of field equations that results from the Poincare group.

But as we have said, the Heisenberg group plays the role of a *translation* group in this non-abelian theory. It should be compared to the translation group $\mathcal{T}(n)$ that is a subgroup of the Poincaré group acting as a dynamical group. It too has trivial Casimir invariants, the generators themselves, and does not lead to particularly interesting field equations.

This suggests that to construct a dynamical group of the form $\mathcal{K} \otimes_s \mathcal{H}(n)$ that generalizes the Poincare group. One can then study the unitary irreducible representations of this group, its Hilbert space and field equations and Casimir invariants and determine whether the resulting theory is of physical interest.

To illustrate how this might work consider $\mathcal{K} = \mathcal{U}(n)$[10] and define $C(n) = \mathcal{U}(n) \otimes_s \mathcal{H}(n)$. The unitary irreducible representations of this group are probably not as familiar as the Poincaré or Heisenberg groups. The Mackey theory is reviewed in the body of the text and for some introductory remarks, we just quote certain results. The generators of $\mathcal{U}(n)$ are $Z_{i,j}$ and the generators of $\mathcal{H}(n)$ may be written $\{A_i^\pm, I\}$ with $A_i^\pm = \frac{1}{\sqrt{2}}(Q_i \pm i P_i)$. The Casimir invariants are given by $\{I, W, W^1, ... W^\alpha, .. W^{N_c}\}$ where $W_{i,j} = A_i^- A_j^+ - I Z_{i,j}$ and

$$W^\alpha = W_{i_1,i_2} W_{i_2,i_3} ...... W_{i_\alpha, i_1}.$$

For the quantum (i.e the representation of $I$ is non-trivial), the little group is $\mathcal{U}(n)$ itself and the Hilbert space for the unitary irreducible representations $\varrho$ of $C(n)$ is $\mathsf{H}^\varrho = \mathfrak{V}^N \otimes \mathfrak{L}^2(\mathbb{R}^n, \mathbb{C})$. $\mathfrak{V}^N$ is a finite dimensional vector space with $N$ equal to the dimension of the unitary irreducible representations of $\mathcal{U}(n)$.

The field equations now are clearly very non-trivial. They are, in fact, quantum harmonic oscillators with unitary spin.

$$\varrho'(W^\alpha)|\psi\rangle = c_\alpha |\psi\rangle, \ |\psi\rangle \in \mathsf{H}^\varrho, \ \alpha = 1, 2, ... N_c$$

The scalar field equation, corresponding to the matrix elements of the representation of $W = P^2 + Q^2 - I U$ with higher order invariants identically zero turns out to be just the time-independent harmonic oscillator

$$\left(\left(i \frac{\partial}{\partial q^i}\right)^2 + q^2 - (2 l + n)\right)\psi_l(q) = 0, \psi_l(q) \in \mathfrak{V}^1 \otimes \mathfrak{L}^2(\mathbb{R}^n, \mathbb{R})$$

where we are using natural units. Note that for the non-quantum limit, the representations degenerate to the representations of the group $\mathcal{U}(n) \otimes \mathcal{T}(2n)$. Here the representation is quite different. In this case, the little group is $\mathcal{U}(n-1)$ and the Hilbert space $\mathsf{H}^\varrho$ is over the projective spheres $\mathbb{P}^{n-1} \simeq \mathcal{U}(n)/\mathcal{U}(n-1)$, $\mathsf{H}^\varrho = \mathfrak{L}^2(\mathbb{P}^{n-1}, \mathbb{R})$.

This is the example mentioned above in the discussion of the Poincaré group. Note that the mathematical dynamical group formalism of the Poincaré group, Heisenberg group and this group $C(n)$ is identical. All lead to field equations for quantum *particle* states, albeit with very different physical interpretations. The interesting point is that the quantum case of the representations $C(n)$ is not constrained to a Hilbert space over a surface in $\mathbb{R}^n$ corresponding to a energy state as in the classical oscillator or a mass state as in the Poincaré case. Rather, the Hilbert space is, in a sense, over all of these *energy states,* that is all of $\mathbb{R}^n$ and it is the field equations define the quantum energy states as is very well know for the quantum oscillator. As these field equations are simply the eigenvalue equation for the representations of the Casimir invariant, they are invariant under the dynamical equation $C(n)$ and, in particular under the non-abelian *translations* generated by $\mathcal{H}(n)$ on the non-abelian quantum position-momentum space.

We are looking for a dynamical group that encompasses both the Heisenberg and the Poincaré groups as dynamical subgroups. We are looking for a group that, in a limit, reduces to the Poincaré group that addresses some of the concerns raised above. Particularly, the Heisenberg non-abelian *translations* should be intrinsic rather than being added through a second quantization prescription. A goal of the theory would also to introduce a generalization of the mass shells analogous to the above discussion of the difference between the classical and quantum oscillators. Clearly the field equations have to be rich enough to encompass the existing equations that arise out of the Poincaré group in an appropriate limit. A very interesting group that exhibits many of these properties is the group $C(1, n) = \mathcal{U}(1, n) \otimes \mathcal{H}(1, n)$, that we name the *quaplectic group*. As we shall shortly describe, it has a number of interesting properties that make it a candidate for study as a dynamical group generalizing and encompassing both the Poincaré and Heisenberg dynamical theories.

Generalizations of this type cannot be deductively obtained. The choice of the more general dynamical group can only motivated by heuristic arguments and guided by principals of simplicity and mathematical elegance. However, once the group is chosen, the results directly follow deductively from the mathematical formalism. No other assumptions are possible or necessary; the Hilbert space of states and the field equations follow directly from the unitary irreducible representations of the group in question. The challenge is that the unitary irreducible representations for a group such as $C(1, n)$ for $n = 3$ are very rich and consequently computational complex. Furthermore, the new effects are expected to be manifest for very strongly interacting systems approaching the Planck scale where the full quantum formalism must be used as a classical approximation is expected to be of little value.

## 1.4 The dynamical group

### 1.4.1 *Classical* motivations

The Hilbert space of the unitary irreducible representations of the quaplectic dynamical group are conjectured to define particle states in a strongly interacting regime analogous to the free particle states of the Poincaré group. The representations of the Casimir operators define field equations for states with physically measurable properties described by the eigenvalues. In the Poincaré case, these are mass and spin (or helicity).

The Poincaré dynamical group has its roots in a *classical* theory. The group acts on the Minkowski space that may be identified with the symmetric space $\mathbb{M}^n = \mathcal{E}(1, n) / \mathcal{SO}(1, n) \simeq \mathbb{R}^n$. This action is not unitary and is finite dimensional. (As we have noted the quantum theory is formulated on Hilbert spaces over hyperboloid surfaces that foliate this space.) The first Casimir invariant is essentially the pseudo-orthogonal metric of this space.

We can construct a *classical* theory of any dynamical group $\mathcal{G} = \mathcal{K} \otimes_s \mathcal{N}$, with $\mathcal{K}, \mathcal{N}$ matrix groups, acting on a symmetric space $\mathbb{A} = \mathcal{G}/\mathcal{K}$. Then a basis of the tangent space $T_a \mathbb{A}$ may be identified with a basis of $\mathfrak{a}(\mathcal{N})$. $\mathcal{N}$ is a normal subgroup and therefore $[K, N] \in \mathfrak{a}(\mathcal{N})$ for all $K \in \mathcal{K}$ and $N \in \mathcal{N}$. The action of $\mathcal{G}$ on $T_a \mathbb{A}$ is given by $g : T_a \mathbb{A} \to T_a \mathbb{A} : N \mapsto g^{-1} N g$ with $g \in \mathcal{G}$ or lifted to the algebra $\mathcal{G}$, the infinitesimal transformations $K : T_a \mathbb{A} \to T_a \mathbb{A} : N \mapsto N + [K, N]$ where $K \in \mathfrak{a}(\mathcal{K})$. This applied to the Poincaré group gives the familiar usual results of basic special relativity.

$$K : T_a \mathbb{A} \to T_a \mathbb{A} : N \mapsto N + [K, N] \tag{1.1}$$

where $K \in \mathfrak{a}(\mathcal{SO}(1, n))$ and $N \in \mathfrak{a}(\mathcal{T91}, n))$. This applied to the Poincaré group, as shown below, gives the familiar usual results of basic special relativity. The Heisenberg case does not appear to have such a classical origin. The space $\mathbb{A} = \mathcal{H}(n) / \mathcal{T}(n)$ has no metric and is not very interesting.

Thus, we have to be careful in looking for a classical motivation of an intrinsically quantum theory. It is not clear that there is an analogue of the quaplectic group in our classical macroscopic experience. Any effects that are in the Planck regime will almost assuredly require a full quantum theory. Never-the-less, any such classical motivation provides an intuitive basis that can help ascribe physical meaning to the theory.

We expect any deep dynamical symmetry to have *remnants* in the classical macroscopic regime. What do we mean by a remnant. A remnants of the Einstein special relativity theory is the Galilei group approximation of nonrelativistic physics that may be obtained from the Poincaré group by simple group contraction. Thus the basic Newtonian concept that velocities add with Euclidean addition is a remnant of the hyperbolic position-space rotations of the Lorentz group. Clearly, the group must contract effectively to the Poincaré group in the low interaction limit. We suggest here the very simple fact that forces add with Euclidean addition is a similar *remnant*. Another interesting question is where does the symplectic group of classical Hamiltonian mechanics come from. We suggest it is a remnant arising as it is precisely the condition required to enable the Heisenberg group to be the normal subgroup of the quaplectic group and therefor physically take on the role corresponding to the *translation* group in this nonabelian quantum theory[10]. These are explored briefly in what follows as heuristic motivation of the full quantum theory.

The quaplectic dynamical group acts in a *classical* theory naturally on the symmetric space $\mathbb{Q}^n \simeq C(1, n)/\mathcal{SU}(1, n)$. This is a ten dimensional non-abelian phase space. It must be emphasized that the classical theory is only of value to give some intuitive insight into the dynamical group in a simple context. These are to motivate the study of the full quantum theory which, as we have noted, has very considerable computational complexity.

The symmetric space $\mathbb{Q}^n$ is non-abelian as the Heisenberg group acts as the *translation* group on this space and the degrees of freedom do not commute. Never-the-less, it is still simply a finite ten dimensional manifold to which usual methods apply. Non manifolds of nonabelian groups are, of course, common in classical theories. A very simple example is the manifold for the rotation group which is nonabelian with noncommuting generators.

1.4.2 Poincaré group

To establish notations and method in a familiar context, consider first the algebra of the Poincaré group. A general element of the Poincaré algebra is $L + A$ where $L \in \mathfrak{a}(\mathcal{SO}(1, n))$ is an element of the Lorentz algebra and $A \in \mathfrak{a}(\mathcal{T}(1, n))$ is an element of the translation algebra.

$$L = \alpha^i J_i + \beta^i K_i, \quad A = \frac{t}{\lambda_t} T + \frac{q^i}{\lambda_q} Q_i \qquad (1.2)$$

The parameters $t$ have the dimensions of time, $\lambda_t$, $q^i$ the dimensions of length $\lambda_q$ and the hyperbolic rotation boost $\beta^i$ and angles $\alpha^i$ are dimensionless. $i, j = 1, , n$.

We digress at this point to discuss dimensional scales. Generally, throughout most of this paper, natural scales for which $c = \hbar = G = 1$ are assumed. We occasionally have reason to introduce explicit scales to study the limiting behavior of the theory. For example, small velocities relative to $c$ are given by the group or algebra contraction with $c \to \infty$.

The usual Planck scales for time, position, momentum and energy dimensioned degrees of freedom are

$$\lambda_t = \sqrt{\hbar/bc}, \lambda_q = \sqrt{\hbar c/b}, \lambda_p = \sqrt{\hbar b/c}, \lambda_e = \sqrt{\hbar b c} \qquad (1.3)$$

where, for convenience we have defined $b = c^4/G$[12]. What are these three basic constants $c, b, \hbar$? Suppose that we measured $x$ direction in meters and $y$ direction in feet. Then, if we believe rotation is a dynamical symmetry, then every time we applied the rotation we would have to insert a constant to convert feet into meters. Equivalently, the units can be made dimensionless by dividing $x$ by feet and $y$ by meters or by choosing natural units where the dimensions are the same. This is all *c* is. Under the Lorentz dynamical transformation it has units of velocity changes meters into seconds. A natural approach is to choose units in which position and time have the same measure by setting $c = 1$. We shall see that under the quaplectic group $\hbar$ and *b* are have a similar character in that they are simply unit conversions for a larger dynamical symmetry acting on the non-abelian phase space of position, time, energy and momentum degrees of freedom.

Looking ahead to the quaplectic group case, *b* translates units of time to units of momentum or units of position to units of energy. It may appear troublesome to be considering a theory where momentum and energy may be mixing with space and time. We have the intuition that the former is a *real* space and the latter is not. However, while it is easy to hold up a hand and point three fingers in orthogonal spacial or *position* directions, it is not possible to point the forth in the direction of *time*. The same is true of pointing in the direction of *energy* and *momentum*. What is important is that the dynamical symmetry acts and *mixes* these degrees of freedom. Furthermore, just as the relativistic effects of time and position mixing are manifest only when velocities approach *c*, these additional affects manifest only when interacting forces between particle states approach *b*, which may be very large. These types of interactions are expected to occur only in regimes where quantum effects are manifest, the forces are defined only between these basic quantum states that are the irreducible unitary representations of the dynamical group.

Returning to the Poincaré algebra, the infinitesimal transformations $\tilde{A} = A + [L, A]$ defined in (1.1) for a basis of the Poincaré algebra and $\tilde{A} = \frac{t}{\lambda_t} \tilde{T} + \frac{q^i}{\lambda_q} \tilde{Q}_i$. Note we choose to transform the basis rather than the parameters. Restricting $n = 3$ for these introductory remarks, the Poincaré algebra is

$$[J_i, J_j] = \epsilon_{i,j}^k J_k, \ [J_i, K_j] = \epsilon_{i,j}^k K_k,$$
$$[K_i, K_j] = -\epsilon_{i,j}^k J_k, \qquad\qquad (1.4)$$
$$[J_i, Q_j] = \epsilon_{i,j}^k Q_k, \ [K_i, Q_j] = c\, \delta_{i,j} T, \ [K_i, T] = \frac{1}{c} Q_i$$

Inserting the Lorentz *L* and translation *A* general elements into (1.1) gives

$$t \tilde{T} + \frac{q^i}{c} \tilde{Q}_i = t T + \frac{q^i}{c} Q_i + \left[\alpha^i J_i + \beta^i K_i, t T + \frac{q^i}{c} Q_i\right]$$

and a straightforward computation with the algebra (1.4) gives

$$\tilde{T} = T + \frac{\beta^i}{c} Q_i$$
$$\tilde{Q}_i = Q_i + \epsilon_{i,j}^k \alpha^i Q_j + c\, \beta^i T$$

The second order Casimir invariant is $C_2 = -\frac{1}{\lambda_t^2}(T^2 - Q^2/c^2)$. Exponentiation to the group using

$$\tilde{A} = e^{\beta^i K_i} A e^{-\beta^i K_i}$$

one obtains the usual finite transformation equations.

$$\begin{aligned}\tilde{T} &= \cosh\beta\, T + \tfrac{\sinh\beta}{\beta}\beta^i Q_i/c,\\ \tilde{Q}_i &= Q_i + \tfrac{\cosh\beta - 1}{\beta^2}\beta^i \beta^j Q_j + \tfrac{\sinh\beta}{\beta} c\beta^i T,\end{aligned} \quad (1.5)$$

where $\beta = \sqrt{\beta^i \beta^i}$. The finite velocity is then (given in one dimension for simplicity) of the form $v = c\tanh\beta$.

In the limit $\beta^i \to 0$ or equivalently $\beta^i = v^i/c$ with $c \to \infty$, the Poincaré group $\mathcal{E}(1,n)$ contracts to the Galelei group $\mathcal{G}a(n+1) \simeq \mathcal{E}(n) \otimes_s \mathcal{T}(1,n)$. $v^i$ now has the dimensions of velocity. In the limit $c \to \infty$ the general element $L$ of the Lorentz group contracts to an element $L°$ of the Euclidean group,

$$L = \alpha^i J_i + \beta^i K_i \to L° = \alpha^i J_i + v^i G_i \quad (1.6)$$

where $G_i = c K_i$ are the translation generators of the Euclidean group $\mathcal{E}(3)$ in $\mathcal{G}a(4)$. Substituting into (1.4)

$$\begin{aligned}[J_i, J_j] &= \epsilon_{i,j}^k J_k, \quad [J_i, G_j] = \epsilon_{i,j}^k G_k, \quad [G_i, G_j] = -\tfrac{1}{c^2}\epsilon_{i,j}^k J_k,\\ [J_i, Q_j] &= \epsilon_{i,j}^k Q_k, \quad [G_i, Q_j] = \delta_{i,j} T, \quad [G_i, T] = \tfrac{1}{c^2} Q_i\end{aligned}$$

and so in the limit $c \to \infty$, the Galelei[11] algebra nonzero commutation relations are

$$\begin{aligned}[J_i, J_j] &= \epsilon_{i,j}^k J_k, \quad [J_i, G_j] = \epsilon_{i,j}^k G_k,\\ [J_i, Q_j] &= \epsilon_{i,j}^k Q_k, \quad [G_i, Q_j] = \delta_{i,j} T\end{aligned} \quad (1.7)$$

The transformation equations are

$$\begin{aligned}\tilde{T} &= T + [L°, T] = T\\ \tilde{Q}_i &= Q_i + [L°, Q_i] = Q_i + \epsilon_{i,j}^k \alpha^i Q_j + v^i T\end{aligned}$$

and the first Casimir invariant of the Galelei group is $C_2 = \frac{1}{\lambda_t^2} T^2$. Exponentiation to the group using

$$\tilde{A} = e^{v^i G_i} A e^{-v^i G_i}$$

one obtains the usual finite transformation equations.

$$\begin{aligned}\tilde{T} &= T,\\ \tilde{Q}_i &= Q_i + v^i T,\end{aligned} \quad (1.8)$$

Now, the Poincaré also acts on the energy-momentum space which may also be, as a manifold, identified with $\mathbb{M}^n$. In this case, $A = \frac{e}{\lambda_e} E + \frac{p^i}{\lambda_p} P_i$

$$\begin{aligned} \tilde{E} &= E + [L, E] = E + c\,\beta^i P_i, \\ \tilde{P}_i &= P + [L, P_i] = P_i + \epsilon^k_{i,j} \alpha^i P_j + \beta^i E/c \end{aligned} \quad (1.9)$$

where now the dimensioned algebra relations are

$$[J_i, P_j] = \epsilon^k_{i,j} P_k, \; [K_i, P_j] = \tfrac{1}{c} \delta_{i,j} E, \; [K_i, E] = -c\, P_i \quad (1.10)$$

and the corresponding first order Casimir invariant is $\tilde{C}_2 = -\frac{1}{\lambda_p^2}(E^2/c^2 - P^2)$. Again with $K_i = c\, G_i$ the Galilean limit of the algebra is again given by (1.7) and the transformation equations are

$$\begin{aligned} \tilde{E} &= E - v^i P_i \\ \tilde{P}_i &= P_i + \epsilon^k_{i,j} \alpha^i P_j \end{aligned}$$

The second order Casimir invariants of the Poincaré algebras have the limiting form $C_2 \to -T^2$ and $\tilde{C}_2 \to P^2$. Note that contracting the Poincaré to the Galelei algebra is a simple deductive process, there is no simple analytic process for the converse. In fact, this converse process took the insight of Einstein. However, once the Poincaré group is identified as the dynamical group, the rest follows directly.

### 1.4.3 The quaplectic group

The above discussion is applicable to any dynamical group that is constructed as a semidirect product. We now turn to the quaplectic group acting on a non-abelian phase space. First, the generators of the Heisenberg algebra are $\{T, Q_i, E, P_i, I\}$. The Heisenberg group plays the role of the *translation* group on the nonabelian phase space. The Lorentz group acting on the position-time and momentum-energy subspaces defines the dynamical group $\mathcal{SO}(1, n) \otimes_s \mathcal{H}(1, n)$. (Note that the group is $\mathcal{SO}(1, n)$ and not $\mathcal{SO}(1, n) \otimes \mathcal{SO}(1, n)$ as the parameters $\alpha^i$, $\beta^i$ are the same for the action on each of these subspaces; it is the same group. Also, $\mathcal{H}(1, n) \simeq \mathcal{H}(1 + n)$[8] is simply a notational convenience.

The dynamical group $\mathcal{SO}(1, n) \otimes_s \mathcal{H}(1, n)$ can be investigated using the mathematical formalism to understand its properties. However, it is not the most general group that can be constructed. It is well known that classical Hamiltonian mechanics on an abelian phase space has a symplectic structure $\mathcal{Sp}(2(n+1))$. This symplectic structure is related to the Heisenberg group acting on a non-abelian phase space in that $\mathcal{Sp}(2n)$ is a subgroup of the group of automorphisms of $\mathcal{H}(n)$. Thus, if we are to construct a semi-direct product of the form $\mathcal{K} \otimes_s \mathcal{H}(1, n)$, $\mathcal{H}(1, n)$ must be normal subgroup. This is only possible if $\mathcal{K}$ is a subgroup of the group automorphisms, and in fact the relevant subgroup of the group of automorphism is the $Sp(2(n+1))$ subgroup[10]. In fact, the condition that the parameters for the action of the Lorentz group acting on the position-time and energy-momentum space are the same is precisely the condition to make the group a subgroup of $Sp(2(n+1))$.

$$Sp(2(n+1)) \cap (SO(1,n) \otimes SO(1,n)) \simeq SO(1,n)$$

This leads us to consider combining the invariants $-\frac{1}{\lambda_t^2}(T^2 - Q^2/c^2)$ and $-\frac{1}{\lambda_p^2}(E^2/c^2 - P^2)$ into the single invariant $\frac{1}{\lambda_t^2}(-T^2 + \frac{1}{c^2}Q^2 + \frac{1}{b^2}(-\frac{1}{c^2}E^2 + P^2))$. The first invariant has dimensions of time squared and the second momentum squared. The constant $b$ has dimensions of force as required. This is the only new physical assumption in this paper and it follows the same idea as was used to generalize from non-relativistic to relativistic mechanics. The maximal invariance group for this invariant is $O(2,6)$. Thus, as the group $\mathcal{K}$ must also be a subgroup of $Sp(2(n+1))$, the homogeneous group is

$$\mathcal{K} = O(2,6) \cap Sp(2(n+1)) = \mathcal{U}(1,n).$$

This gives us the hypothesis of the quaplectic group

$$C(1,n) = \mathcal{U}(1,n) \otimes_s \mathcal{H}(1,n) \simeq S\mathcal{U}(1,n) \otimes_s (Os(1,n))$$

where

$$Os(1,n) = \mathcal{U}(1) \otimes_s \mathcal{H}(1,n)$$

Now

$$-T^2 + \tfrac{1}{c^2} Q^2 + \tfrac{1}{b^2} P^2 - \tfrac{1}{b^2 c^2} E^2 \tag{1.11}$$

is an invariant of $\mathcal{U}(1,n)$ but it is not an invariant of the full group $C(1,n)$. Another term $\frac{2}{bc} U I$ must be added that is also $\mathcal{U}(1,n)$ invariant to make the expression invariant under the nonabelian *translations* $\mathcal{H}(1,n)$.

$$C_2 = -\tfrac{1}{2\lambda_t^2}(T^2 - \tfrac{1}{c^2}Q^2 - \tfrac{1}{b^2}P^2 + \tfrac{1}{b^2 c^2}E^2) - UI \tag{1.12}$$

where $U$ is the generator of the $\mathcal{U}(1)$ algebra and $I$ is the center of the Heisenberg algebra. A general dimensioned element $Z$ of the algebra of $\mathcal{U}(1,n)$ and $A$ of the algebra of the Heisenberg group $\mathcal{H}(1,n)$ may be written as

$$\begin{aligned} Z &= \beta^i K_i + \gamma^i N_i + \alpha^i J_i + \theta^{i,j} M_{i,j} + \vartheta R, \\ A &= \tfrac{1}{\lambda_t}\left(tT + \tfrac{e}{cb}E + \tfrac{q^i}{c}Q_i + \tfrac{p^i}{b}P_i\right) + \iota I. \end{aligned} \tag{1.13}$$

Note that the generators $\{J_i, K_i, Q_i, T\}$ define the Poincaré algebra of the position-time subspace given above in (1.4) and that the generators $\{J_i, K_i, P_i, E\}$ define the Poincaré algebra of the momentum-energy subspace given above in (1.10).

The generator $U$ of the $\mathcal{U}(1)$ algebra in the factorization $\mathcal{U}(1, n) \simeq \mathcal{U}(1) \otimes \mathcal{SU}(1, n)$ given above is

$$U = \tfrac{1}{2} \left( -R + \delta^{i,j} M_{i,j} \right)$$

A new, and interesting feature is that the generators $\{J_i, N_i, P_i, T\}$ also define a Poincaré algebra of the momentum-time subspace.

$$[J_i, J_j] = \epsilon_{i,j}^k J_k, \quad [J_i, N_j] = \epsilon_{i,j}^k N_k, \quad [N_i, N_j] = -\epsilon_{i,j}^k J_k,$$
$$[J_i, P_j] = \epsilon_{i,j}^k P_k, \quad [N_i, P_j] = b \, \delta_{i,j} T, \quad [N_i, T] = \tfrac{1}{b} P_i$$

and $\{J_i, N_i, Q_i, E\}$ with the translation subspace generators now. The finite transformations using

$$\tilde{A} = e^{\gamma^i N_i} A \, e^{-\gamma^i N_i}$$

are the finite transformations for force boosts that correspond to the usual velocity boost equations given in (1.14)

$$\begin{aligned}
\tilde{T} &= \cosh \gamma \, T + \tfrac{\sinh \gamma}{\gamma} \gamma^i P_i / b, \\
\tilde{P}_i &= P_i + \tfrac{\cosh \gamma - 1}{\gamma^2} \gamma^i \gamma^j P_j + \tfrac{\sinh \gamma}{\gamma} b \gamma^i T,
\end{aligned} \qquad (1.14)$$

where $\gamma = \sqrt{\gamma^i \gamma^i}$. The finite force is then (given in one dimension for simplicity) of the form $f = b \tanh \gamma$.

$$[J_i, J_j] = \epsilon_{i,j}^k J_k, \quad [J_i, N_j] = \epsilon_{i,j}^k N_k, \quad [N_i, N_j] = -\epsilon_{i,j}^k J_k,$$
$$[J_i, Q_j] = \epsilon_{i,j}^k Q_k, \quad [N_i, Q_j] = \tfrac{1}{b} \delta_{i,j} E, \quad [N_i, E] = -b Q_i$$

This means that rates of change of momentum, that is, force, is bounded. Furthermore, it means that the concept of force relative to *empty space* does not have meaning, it can only be relative to particle states that are irreducible unitary representations of the dynamical group. Note that in the small interaction limit given by the limit $c, b \to \infty$ gives the usual Euclidean addition law for forces of non-relativistic Hamiltonian mechanics.

The nonzero $R$ generators are

$$[K_i, R] = N_i, \; [N_i, R] = K_i \; [R, T] = \tfrac{1}{bc} E, \; [R, E] = -b c T$$

The discussion of the remaining commutators involving $M_{i,j}$ are deferred to the body of the text (3.13).

The generators $\{T, E, Q_i, P_i, I\}$ define the Heisenberg algebra

$$[Q_i, P_j] = \hbar \, \delta_{i,j} I, \; [T, E] = -\hbar I, \qquad (1.15)$$

where we have noted that $\lambda_t \lambda_e = \lambda_q \lambda_p = \hbar$. Also, it may appear that we are missing a factor of $i$. This is the algebra of the Heisenberg group itself, the $i$ appears from the Hermitian representation of the quantum theory[3]. Note that these are compatible with the above Poincaré algebras as only the generators $\{T, Q_i\}$, $\{E, P_i\}$, $\{T, P_i\}$, $\{E, Q_i\}$ spanning the $\mathcal{T}(1, n)$ subgroups of $\mathcal{H}(1, n)$ appear.

A useful diagram for thinking about this non-abelian phase space is the diagram

$$\begin{array}{ccc} T & \leftrightarrow & Q_i \\ \updownarrow & & \updownarrow \\ P_i & \leftrightarrow & E \end{array}$$

Only generators appearing on an edge can be simultaneously diagonalized. We say a formulation in which $\{T, Q_i\}$ is diagonal is *orthogonal* to $\{T, P_i\}$ in this sense. Note that the dimensionality of the basic scales may be illustrated in a similar quad diagram. That is the physical dimension of $Q_i/T$ and $E/P_i$ is $c$ and $P_i/T$ and $E/Q_i$ is $b$ and $Q_i P_i$ and $E T$ is $\hbar$.

$$\begin{array}{ccccc} \lambda_t & - & c = \lambda_q/\lambda_t & - & \lambda_q \\ | & \nwarrow & & \nearrow & | \\ b = \lambda_p/\lambda_t & & \hbar = \lambda_q \lambda_p = \lambda_t \lambda_e & & b = \lambda_e/\lambda_q \\ | & \swarrow & & \searrow & | \\ \lambda_p & - & c = \lambda_e/\lambda_p & - & \lambda_e \end{array}$$

The combining of the invariants has introduced an *orthogonal* relativity principal. That is, it is a bound of the relative forces that may be experienced in the theory by the constant $b$. Note that in the general quantum theory where we consider the unitary irreducible representations of these generators acting on a Hilbert space of states, one may only diagonalize the translation generators for one of the above four Poincaré algebras. That is, only one of these *relativity principles* is observable at a given time.

The general transformation equations $A' = A + [Z, A]$ are

$$\begin{aligned} \tilde{T} &= T + \beta^i Q_i/c + \gamma^i P_i/b + \vartheta E/cb, \\ \tilde{E} &= E - c\gamma^i Q_i + b\beta^i P_i - bc\vartheta T, \\ \tilde{Q}_i &= Q_i + \epsilon_{i,j}^k \alpha^i Q_j + c\beta^i T - \gamma^i E/b + c\theta^{i,j} P_j/b, \\ \tilde{P}_i &= P_i + \epsilon_{i,j}^k \alpha^i P_j + \beta^i E/c + b\gamma^i T - b\theta^{i,j} Q_j/c, . \end{aligned} \quad (1.16)$$

In these equations, all of the phase space degrees of freedom *mix*. However, this *mixing* only is significant when $\beta^i$, $\gamma^i$, $\vartheta$ and $\theta^{i,j}$ are large. Note that exponentiation to the group for just the boost generators using

$$\tilde{A} = e^{\beta^i K_i + \gamma^i N_i} A\, e^{-\beta^i K_i - \gamma^i N_i}$$

one obtains the finite transformation equations. These are the generalizations of (1.5) and (1.14) in this theory.

$$\begin{aligned}
\tilde{T} &= \cosh\zeta\, T + \tfrac{\sinh\zeta}{\zeta}(\beta^i Q_i/c + \gamma^i P_i/b), \\
\tilde{E} &= \cosh\zeta\, E + \tfrac{\sinh\zeta}{\zeta}(-b\,\gamma^i Q_i + c\,\beta^i P_i), \\
\tilde{Q}_i &= Q_i + \tfrac{\cosh\zeta-1}{\zeta^2}\zeta^{i,j} Q_j + \tfrac{\sinh\zeta}{\zeta}(c\,\beta^i T - \gamma^i E_i/b), \\
\tilde{P}_i &= P_i + \tfrac{\cosh\zeta-1}{\zeta^2}\zeta^{i,j} P_j + \tfrac{\sinh\zeta}{\zeta}(b\,\gamma^i T + \beta^i E/c).
\end{aligned} \quad (1.17)$$

where $\zeta^{i,j} = \beta^i\beta^j + \gamma^i\gamma^j$ and $\zeta = \sqrt{\delta_{i,j}\zeta^{i,j}}$.

Now, the limit corresponding to the Galelei limit of the Poincaré group is now $b, c \to \infty$. As with the Poincare case, the parameters are scaled with the dimensional constants.

$$\beta^i = \tfrac{v^i}{c},\; \gamma^i = \tfrac{f^i}{b},\; \vartheta = \tfrac{1}{bc}r^\circ,\; \theta^{i,j} = \tfrac{r^{i,j}}{bc}$$

$f^i$ has the dimensions of force and $r^{i,j}$ and $r^\circ$ power. As $b, c \to \infty$, these parameters go to zero. The corresponding scaling of the generators are

$$N_i = b\,F_i,\; K_i = c\,G_i,\; R = b\,c\,R^\circ,\; M_{i,j} = b\,c\,M^\circ{}_{i,j}$$

The nonzero generators of the limiting form of the algebra are the Heisenberg relations [1.15] and

$$[J_i, J_j] = \epsilon^k_{i,j} J_k, \quad [R^\circ, E] = T,$$

$$\begin{aligned}
&[J_i, G_j] = \epsilon^k_{i,j} G_k,\; [J_i, P_j] = \epsilon^k_{i,j} P_k,\; [G_i, E] = P_i,\; [G_i, Q_i] = \delta_{i,j} T \\
&[J_i, F_j] = \epsilon^k_{i,j} F_k,\; [J_i, Q_j] = \epsilon^k_{i,j} Q_k,\; [F_i, E] = Q_i,\; [F_i, P_j] = \delta_{i,j} T
\end{aligned} \quad (1.18)$$

$$[G_i, F_j] = M^\circ{}_{i,j} + \delta_{i,j} R^\circ,\quad [Q_i, P_j] = \hbar\,\delta_{i,j} I,\; [T, E] = -\hbar I,$$

Note that the commutators of the $M^\circ{}_{i,j}$ are all identically zero and thus these generators are all trivial. The general element of the contracted group is

$$Z^\circ = v^i G_i + f^i N_i + \alpha^i J_i + r^\circ R^\circ,$$

The corresponding transformation equations are what one expects for a classical phase space. Here, in addition to the usual velocity $v^i$ the parameter $f^i$, is identified with the classical force that satisfies the usual Euclidean addition law. Inspection of the remaining commutation relations is what one expects in the nonrelativistic limit and the corresponding transformations are

$$\begin{aligned}
\tilde{T} &= T, \\
\tilde{E} &= E - \gamma^i Q_i + \beta^i P_i - r^\circ T,
\end{aligned}$$

$$\tilde{Q}_i = Q_i + \epsilon^k_{i,j}\,\alpha^i\,Q_j + \beta^i\,T,$$
$$\tilde{P}_i = P_i + \epsilon^k_{i,j}\,\alpha^i\,P_j + \gamma^i\,T$$

Finally, the contracted form of the finite boost equations given by

$$\tilde{A} = e^{v^i\,G_i + f^i\,F_i}\,A\,e^{-v^i\,G_i - f^i\,F_i}$$

are what you would expect from elementary nonrelativistic mechanics

$$\begin{aligned}\tilde{T} &= T\,,\\ \tilde{E} &= E - f^i\,Q_i + v^i\,P_i,\\ \tilde{Q}_i &= Q_i + v^i\,T,\\ \tilde{P}_i &= P_i + f^i\,T.\end{aligned} \quad (1.20)$$

In the general equations, all of the degrees of freedom are on equal footing in the non-abelian phase space.  For some reason, this often seems alien and so a few heuristic remarks are warranted.  We have a deeply ingrained notion of the four dimensional position-time manifold.  We are very comfortable stating that *time* is the *same* as *position.* In the sense of the Poincaré group acting on it as a dynamical group this is true. As we have noted constant $c$ is regarded simply as a conversion of dimensions of seconds into dimensions of meters and $b$ converts dimensions of seconds into momentum units $kg\,m/s$ as we hyperbolically rotate time into position by the generators $K_i$ and $N_i$.  Choosing natural units is equivalent puts everything in the same dimensional units so these conversion constants are not required.  The fact remains that we can point three fingers in the different orthogonal position dimensions but cannot literally point the fourth in the direction of time.  The situation on the full phase space is no different. The quaplectic dynamical group transforms the degrees of freedom into each other.  Again, one cannot directly point in the direction of  energy or momentum any more than one can point in the direction of time. However, in a theory with the quaplectic dynamical group, they are no less real or physical than the underlying position and time degrees of freedom. The situation is made more challenging by the fact that in the quantum phase space is nonabelian.

1.4.4  Born reciprocity in non-relativistic classical and quantum mechanics

It is of value to show how this *orthogonal* momentum-time   representation is present even in the most elementary non-relativistic classical  mechanics. This digression is to again show that the bias to the position time formulation of mechanics is a predilection that is not completely demanded by the formalism.

Hamiltonian mechanics, the equations of motion are invariant under the transform

$$\frac{dq^i}{dt} = \frac{\partial H(q^i,p^i)}{\partial p^i},\quad \frac{dp^i}{dt} = -\frac{\partial H(q^i,p^i)}{\partial q^i}$$

are invariant under the Born reciprocity transform [7,8]

$$\{t, e, q, p\} \rightarrow \{t, e, p, -q\},$$

The Lagrangian formulation obtained through the Legendre transformation

$$L(q^i, \dot{q}^i) = H(q^i, p^i) - \dot{q}^i p^i, \quad \dot{q}^i \doteq \frac{dq^i}{dt}$$

assuming the Hessian is non zero appears to single out the position-time representation as preferred until one realizes that one can equally well Legendre transform in the opposite direction

$$L(p^i, \dot{p}^i) = H(q^i, p^i) + \dot{p}^i q^i, \quad \dot{p}^i \doteq \frac{dp^i}{dt}$$

again with the assumption that the appropriate Hessian does not vanish. Now, the Hessian does vanish for a free particle, but we must remember that the truly free particle is a theoretical abstraction. Then just as the variational principle $\delta \int L(q^i, \dot{q}^i) dt = 0$ is used to derive the Lagrangian equations,

$$0 = \frac{\partial L(q^i, \dot{q}^i)}{\partial q^i} - \frac{d}{dt} \frac{\partial L(q^i, \dot{q}^i)}{\partial \dot{q}^i},$$

a corresponding variational principal

$$0 = \frac{\partial L(q^i, \dot{q}^i)}{\partial q^i} - \frac{d}{dt} \frac{\partial L(q^i, \dot{q}^i)}{\partial \dot{q}^i},$$

that can be also derived from the *orthogonal* variational principle $\delta \int L(p^i, \dot{p}^i) dt = 0$

$$0 = \frac{\partial L(p^i, \dot{p}^i)}{\partial p^i} - \frac{d}{dt} \frac{\partial L(p^i, \dot{p}^i)}{\partial \dot{p}^i},$$

The above properties of the Hamiltonian formulation are inherited by the nonrelativistic quantum mechanics as formulated in Dirac's transformation theory. There is no preference in this theory for a basis in which momentum or position is diagonalized.

More generally, the Poincaré-Cartan form

$$-de \wedge dt + \delta_{i,j} dp^i \wedge dq^j$$

has four canonical one forms

$$-e\, dt + p \cdot dq, \; -e\, dt - q \cdot dp, \; t\, de + p \cdot dq, \; t\, de - q \cdot dp$$

each with a Lagrangian formulation. If the topology is trivial, they are related by the obvious exact forms.

The relativistic theory based on the Poincaré dynamical group do not have this property. However, the quaplectic group re-instates this equivalence of the position and momentum degrees of freedom and adds a similar equivalence of the time and energy degrees of freedom combined in a single non-abelian manifold. Note that the transformation equations in (1.16) are invariant under the transforms.

$$\{t, e, q, p\} \to \{t, e, p, -q\}, \quad \{t, e, q, p\} \to \{e, -t, q, p\}$$

These transforms are the basis of the Born reciprocity principle.

## 2: Mackey theory and the Poincaré and Heisenberg dynamical groups

The quantum theory for the dynamical group requires the unitary irreducible representations to be determined and for the eigenvalue equations for the representations of the Casimir operators to be solved.

The restriction of the class of groups to matrix groups that are algebraic is sufficient to satisfy the conditions for the Mackey unitary irreducible representation theory of semidirect products to apply.

The Heisenberg group has a trivial Casimir operator $I$. The $3+1$ dimensional Poincaré group has two Casimir operators and the associated eigenvalue equations for the invariants define basic equations of physics [13,46]. We review the representation theory and Casimir equations in the current context to establish the connection with the current notation.

The same formalism may be used to study an example theory based on $C(n) = \mathcal{U}(n) \otimes_s \mathcal{H}(n)$. As we are ultimately interested in $C(1, n)$, this is analogous to studying the Euclidean subgroup $\mathcal{E}(n)$ of the Poincaré theory $\mathcal{E}(1, n)$. This system has $n+1$ Casimir invariants that provides a rich set of field equations. This case enables the effects of the non-abelian $\mathcal{H}(n)$ to be illustrated in the compact case $\mathcal{U}(n)$ that has finite dimensional unitary representations. The classical limit to the form $\mathcal{U}(n) \otimes_s \mathcal{T}(2n)$ is briefly discussed.

Finally, the formalism is used to study the canonically relativistic theory based on $C(1, n) = \mathcal{U}(1, n) \otimes_s \mathcal{H}(1, n)$. This system has $n+2$ Casimir invariants that provides a rich set of field equations to study.

### 2.1 Mackey representation theory

The groups of interest are semi-direct products of the form $\mathcal{G} = \mathcal{K} \otimes_s \mathcal{N}$ where $\mathcal{G}, \mathcal{K}, \mathcal{N}$ are matrix groups that are algebraic[5]. The problem is to determine the unitary dual $\hat{\mathcal{G}}$ that is defined to be the complete set of equivalence classes of irreducible unitary representations $\varrho \in \hat{\mathcal{G}}$ and the corresponding representation Hilbert space $\mathsf{H}^\varrho$. The representations $\varrho$ evaluated at a point $g \in \mathcal{G}$ are unitary operators on the Hilbert space $\mathsf{H}^\varrho$ with elements $|\psi\rangle \in \mathsf{H}^\varrho$.

$$\varrho(g) : \mathsf{H}^\varrho \to \mathsf{H}^\varrho : |\psi\rangle \mapsto |\tilde{\psi}\rangle = \varrho(g) |\psi\rangle$$

Mackey [29] showed that the representations of $\mathcal{G}$ may be constructed from the representations of the stabilizer group $\mathcal{G}^\circ = \mathcal{K}^\circ \otimes_s \mathcal{N}$ determined by certain fixed point properties in the natural action of $\mathcal{G}$ on the unitary dual $\hat{\mathcal{G}}$. This applies to a general class of semi-direct product groups subject to certain technical constraints. As noted, a sufficient condition for the theory to apply are that the groups concerned are matrix groups that are algebraic [49]. In a previous paper [28], the calculation of the representations was presented for all the groups of interest in this paper and so this is not repeated here. In this section, we give only the form of the results and notation necessary for the discussions that follow. Our particular emphasis will be on the representations $\varrho'$ lifted to the algebra $\mathfrak{a}(\mathcal{G})$. Additional remarks on the Mackey theory are given in references [28,49,29,32].

Then, $\hat{\mathcal{G}}$ (the space of equivalence classes of unitary irreducible representations) is induced from the representations $\varrho^\circ = \sigma \otimes \rho$ acting on $\mathsf{H}^{\varrho^\circ} = \mathsf{H}^\sigma \otimes \mathsf{H}^\xi$. $\sigma$ are the unitary irreducible representations of the little group $\mathcal{K}^\circ$ on the corresponding Hilbert space $\mathsf{H}^\sigma$. $\xi$ are the unitary irreducible representations of the normal subgroup $\mathcal{N}$ on the Hilbert space $\mathsf{H}^\xi$. $\rho$ are unitary projective representations of the full stabilizer $\mathcal{G}^\circ$ on the Hilbert space $\mathsf{H}^\xi$ associated with the representations $\xi$. $\rho$ are a certain projective extension of the representations $\xi$ to the full stabilizer, $\rho|_\mathcal{N} = \xi$ as described in the references above. If $\mathcal{N}$ is abelian, the extension is trivial and $\rho \simeq \xi$.

The representations $\varrho$ of the full group $\mathcal{G}$ are induced, using the Mackey induction theorem, from the representation $\varrho^\circ$ of the stabilizer subgroup $\mathcal{G}^\circ$. These representations $\varrho$ act on the Hilbert space $\mathsf{H}^\varrho \simeq \mathsf{H}^\sigma \otimes \mathcal{L}^2(\mathbb{A}, \mathbb{R})$ where $\mathbb{A}$ is the symmetric space $\mathbb{A} = \mathcal{G}/\mathcal{G}^\circ$.

The above remarks may be lifted to the algebra, $X \in \mathfrak{a}(\mathcal{G})$. That is, the representations of the algebra $\varrho' = T_e \varrho$ acting on the algebra define Hermitian[3] operators

$$\varrho'(X) : \mathsf{H}^\varrho \to \mathsf{H}^\varrho : |\psi\rangle \mapsto |\tilde{\psi}\rangle = \varrho'(X)|\psi\rangle$$

The little group representation $\varrho^{\circ\prime} = \sigma' \oplus \rho'$ acts on elements $X^\circ \in \mathfrak{a}(\mathcal{G}^\circ)$ and the discussion follows as above. The condition on the extension $\rho$ in terms of its action on the algebra is as follows. Let $\{K_\alpha\}$, $\alpha, \beta = 1, ..\dim(\mathcal{K})$ be a basis of $\mathfrak{a}(\mathcal{K})$ and $\{N_a\}$, $a, b = 1..\dim(\mathcal{N})$ be a basis of $\mathfrak{a}(\mathcal{N})$. Then the representation of the algebra is projective in the sense that

$$\begin{aligned}&\varrho'([N_a, N_b]) = i\, c^c_{a,b}\, \varrho'(N_c) \\ &\varrho'([K_\alpha, N_a]) = i\, s\, c^b_{\alpha,a}\, \varrho'(N_b) \\ &\varrho'([K_\alpha, K_\beta]) = i\, s\, c^\kappa_{\alpha,\beta}\, \varrho'(K_\kappa)\end{aligned} \qquad (2.1)$$

where $s \in \mathbb{R}$ is a real, nonzero constant defining the projective representation and the $c_{a,b}^{c}$, $c_{\alpha,a}^{\beta}$, $c_{\alpha,\beta}^{\kappa}$ are the usual structure constants of the algebra $\mathfrak{a}(\mathcal{G})$. Again, the $i$ appears in the bracket relations as these are Hermitian, not anti-Hermitian representations of the algebra[3], a point that we shall not mention again. As $\varrho'(K_\alpha)$ acts on the Hilbert space $\mathsf{H}^\xi$ of the representations $\xi$ of $\mathcal{N}$, $\varrho'(K_\alpha)$ must be in the enveloping algebra of the representations and therefore

$$\varrho'(K_\alpha) = d_\alpha^a \xi'(N_a) + d_\alpha^{a,b} \xi'(N_a)\xi'(N_a) + .... \qquad (2.2)$$

for some constants $d_\alpha^a$, $d_\alpha^{a,b}$, ... .. Substituting into the above algebra relations defines a system of equations that may be solved for the $d_\alpha^a$, $d_\alpha^{a,b}$, ... .

## 2.2 Poincaré representation theory

### 2.2.1 Basic properties of the algebra

The $n + 1$ dimensional Poincaré group may be taken to be the universal cover $\mathcal{E}(1, n)$. The Lie algebra $\mathfrak{a}(\mathcal{E}(1, n)))$ of the pseudo-Euclidean groups $\mathcal{E}(1, n)$ is spanned by the generators $\{L_{a,b}, Y_c\}$ satisfying the commutators.

$$[L_{a,b}, L_{c,d}] = \eta_{a,d} L_{b,c} + \eta_{b,c} L_{a,d} - \eta_{a,c} L_{b,d} - \eta_{b,d} L_{a,c}$$
$$[L_{a,b}, Y_c] = \eta_{a,c} Y_b - \eta_{b,c} Y_a$$

with $a, b, .. = 0, 1..n$. The rank of the group, corresponding to the number of Casimir invariants $N_c$, is $\frac{n+1}{2}$ rounded up to the nearest integer. The first Casimir invariant is

$$C_2 = Y^2 = (Y, Y) \qquad (2.3)$$

where $(X, Y) \doteq \eta^{a,b} X_a Y_b$. Higher order invariants have complicated expressions [10,38] and for the illustrative purposes here, we just note that for the $n = 2, 3$ cases

$$\begin{aligned} n = 2: C_4 &= \epsilon^{i,j,k} Y_i L_{j,k}, \quad i, j, .. = 1, 2, 3 \\ n = 3: C_4 &= W^2 = (W, W), \quad W^a \doteq \epsilon^{a,b,c,d} Y_b L_{c,d}, \quad a, b, .. = 0, 1, 2, 3 \end{aligned} \qquad (2.4)$$

The rank of $\mathcal{E}(1, 1)$ is 1 as the little group is trivial and so it only has a second order Casimir invariant $C_2$.

For $n = 3$, these may be written in three notation with $K_i = L_{0,i}$ and $J_i = \epsilon_i^{k,j} L_{k,j}$ with $i, j = 1, 2, 3$ and where we are assuming natural units. The algebra is then given by (1.4). It is also useful to define $J^\pm = J_1 \pm i J_2$ with corresponding algebra for the rotation subgroup given by

$$[J^\pm, J_3] = \pm J^\pm, \quad [J^+, J^-] = 2 J_3$$

2.2.2 Unitary irreducible representations and Hilbert space: Mackey theory

Suppose that $\varrho$ are the unitary irreducible representations and that the lift to the algebra is $\varrho'$. The action of the first Casimir invariant on a state $|\psi\rangle \in \mathsf{H}^\varrho$ is

$$\varrho'(C_2)|\psi\rangle \doteq \varrho'(Y^2)|\psi\rangle = c_2 |\psi\rangle$$

As the representation is Hermitian, the eigenvalues are real and the indefinite metric means that $c_2$ can range over the full values of $\mathbb{R}$.

The Mackey representation theory can be applied noting that this is the simpler case for which the normal subgroup is abelian. This representation theory [48] is summarized in the Mackey theory context in [27] and the resulting little groups and corresponding stabilizer groups are as follows. As the stabilizer groups are all proper subgroups of $\mathcal{E}(1, n)$, the Mackey induction theorem must be used. This requires the construction of the symmetric spaces $\mathbb{A}$ that are the quotient of $\mathcal{E}(1, n)$ and the stabilizer groups.

|  | Little Group | Stabilizer Group | Symmetric Space ($\mathbb{A}$) |
|---|---|---|---|
| $c_2 < 0$ | $\mathcal{SO}(n)$ | $\mathcal{E}(n)$ | $\mathbb{H}^n_- \simeq \mathcal{E}(1, n)/\mathcal{E}(n) \simeq \mathcal{SO}(1, n)/\mathcal{SO}(n)$ |
| $c_2 = 0$ | $\mathcal{E}(n-1)$ | $\mathcal{G}a(n)$ | $\mathbb{H}^n_0 \simeq \mathcal{E}(1, n)/\mathcal{G}a(n) \simeq \mathcal{SO}(1, n)/\mathcal{E}(n-1)$ |
| $c_2 > 0$ | $\mathcal{SO}(1, n-1)$ | $\mathcal{E}(1, n-1)$ | $\mathbb{H}^n_+ \simeq \mathcal{E}(1, n)/\mathcal{E}(1, n-1) \simeq \mathcal{SO}(1, n)/\mathcal{SO}(1, n-1)$ |

In the above we have noted that the group $\mathcal{G}a(n) \doteq \mathcal{E}(n-1) \otimes_s \mathcal{T}(n)$, which may be recognized as the Galilei group. In addition, there is the trivial and the degenerative representation for which $\varrho'(Y_a)|\psi\rangle = 0$ and the representation is that of $\mathcal{SO}(1, n)$.

The corresponding unitary irreducible representations $\varrho^{\pm,0}$ of $\mathcal{E}(1, n)$ act on the Hilbert space .

$$\mathsf{H}^{\varrho^{\pm,0}} \simeq \mathsf{H}^{\sigma^{\pm,0}} \otimes \mathcal{L}^2(\mathbb{H}^n_{\pm,0}, \mathbb{R})$$

$\mathsf{H}^{\sigma^{\pm,0}}$ are the Hilbert spaces of the representations of the little groups. For the case $c_2 < 0$ the little group is $\mathcal{SO}(n)$ and the unitary irreducible representations are on the finite dimensional vector space $\mathsf{H}^{\sigma^-} \simeq \mathfrak{V}^M$ where $M = \dim \sigma^-$ is the dimension of these representations. For the null case, the representations of $\mathcal{E}(n-1)$ are generally infinite dimensional. Generally, only the degenerate representations that are equivalent to the finite dimensional unitary irreducible representations of $\mathcal{SO}(n-1)$ are considered to be physical; $\mathsf{H}^{\sigma^0} \simeq \mathfrak{V}^{N^\circ}$ where here $M^\circ = \dim \sigma^0$. The spacelike case with countably infinite dimensional representations of the noncompact $\mathcal{SO}(1, n-1)$ are generally considered not to be physical.

As an aside, note that the little group of the Euclidean group $\mathcal{E}(n)$ are the groups $\mathcal{SO}(n-1)$ and the stabilizer is $\mathcal{E}(n-1)$. The symmetric spaces are the $n$ spheres

$$\mathbb{S}^{n-1} \simeq \mathcal{E}(n)/\mathcal{E}(n-1) \simeq \mathcal{SO}(n)/\mathcal{SO}(n-1)$$

The nondegenerate representation Hilbert space is

$$\mathsf{H}^{\varrho} \simeq \mathfrak{V}^M \otimes \mathcal{L}^2(\mathbb{S}^{n-1}, \mathbb{R}) \text{ with } M = \dim \sigma$$

where, in this context, $\sigma$ are the representations of the little group $\mathcal{SO}(n-1)$. The Casimir operator $C_2$ of $\mathcal{E}(n)$ has eigenvalues $c_2$ that are positive.

Returning to the Poincaré group, consider now the time-like representation $\varrho^-$ acting on $\mathsf{H}^{\varrho^-} \simeq \mathfrak{V}^M \otimes \mathcal{L}^2(\mathbb{H}^n_-, \mathbb{R})$. A basis of the Hilbert space $\mathsf{H}^{\varrho^-}$ is

$$|m\rangle \otimes |\delta_x\rangle \doteq |\delta_x, m\rangle$$

The inner product on the Hilbert space $\mathcal{L}^2(\mathbb{H}^n_-, \mathbb{R})$ is [37]

$$\langle \delta_y \mid \delta_x \rangle \doteq \delta_{x,y} \, \delta_{(x,y), c_2}$$

Then $|\psi\rangle \in \mathsf{H}^{\varrho^-}$ is represented in the basis as $\langle \delta_x, m \mid \psi \rangle = \psi_m(x)$. The $m$ label the $M$ degrees of freedom in $\mathfrak{V}^M$ that are the *internal* spin degrees of freedom and the $\psi(x)$ are square integrable functions on the hyperboloid $\mathbb{H}^{n-1}_-$. Finally, the inner product is then

$$\langle \varphi \mid \psi \rangle = \langle \varphi \mid \delta_x, m \rangle \langle \delta_x, m \mid \psi \rangle = \sum_m \int d^n x \, \delta(\eta_{a,b} x^a x^b - c_2) \, \varphi^\dagger_m(x) \psi_m(x)$$

Similar discussions follow for the Hilbert spaces $\mathcal{L}^2(\mathbb{H}^n_0, \mathbb{R})$ and $\mathcal{L}^2(\mathbb{H}^n_+, \mathbb{R})$ and also for the Hilbert space $\mathcal{L}^2(\mathbb{S}^{n-1}, \mathbb{R})$ associated with the representations of $\mathcal{E}(n)$.

2.2.3 Unitary irreducible representations of the algebra

The representation may be lifted to the tangent spaces to define the representation $\varrho'$ of the algebra. This representation may be decomposed as $\varrho' = \sigma' \oplus \rho'$. $\sigma'$ is the irreducible Hermitian representation of the generators of the little group that acts on the $\mathsf{H}^{\sigma^{\pm,0}}$ degrees of freedom. For the time-like case $\mathsf{H}^{\sigma^-}$, these are the familiar generators of the compact little group $\mathcal{SO}(n)$. For simplicity, we now consider just the case $n = 3$ and then for the little group $\mathcal{SO}(3)$, the unitary irreducible representations of $\mathcal{SO}(3)$ are given by the $2j+1$ dimensional matrices that are the matrix elements

$$\begin{aligned}\Sigma^{\pm}_{m',m} &= \langle m' \mid \sigma'(J^\pm) \mid m \rangle = \sqrt{j(j+1) - m(m \pm 1)} \, \delta_{m', m \pm 1} \\ \Sigma^3_{m',m} &= \langle m' \mid \sigma'(J_3) \mid m \rangle = m \, \delta_{m',m}\end{aligned} \quad (2.5)$$

with $m = -j, \ldots 0, \ldots j$ and $|m\rangle \in \mathsf{H}^{\sigma^-} \simeq \mathfrak{V}^M$ with $M = 2j+1$. Note that $\sigma'(K_i) = 0$ and the representation of generators of the normal subgroup are also zero, $\sigma'(E) = \sigma'(P_i) = 0$

The representations $\rho'$ from the Mackey representations are, in general, the extension of the representations of the normal subgroup to the full group. However, as the normal subgroup is abelian, the extension is trivial and the representation of the generators of the $\mathcal{SO}(1, n)$ subgroup are trivial. Restricting to the case $n = 3$,

$$\rho'(J^\pm)|m\rangle = 0, \ \rho'(J_3)|m\rangle = 0, \ \rho'(K_i)|m\rangle = 0, \ i = 1, 2, 3$$

The representation acting on the algebra of the normal subgroup reduces to $\rho' = \xi'$ where $\xi$ is the unitary irreducible representations of the translation group $\mathcal{T}(1, n)$. These representations on the Hilbert spaces $\mathsf{H}^{\xi^{\pm,0}} \simeq \mathcal{L}^2(\mathbb{H}^n_{\pm,0}, \mathbb{R})$ are given by

$$\langle \delta_x | \xi'(P_i) | \psi \rangle = -i \frac{\partial}{\partial q^i} \psi(q^i, t), \quad \langle \delta_x | \xi'(E) | \psi \rangle = i \frac{\partial}{\partial t} \psi(q^i, t)$$

Putting it together, if $Z$ is an arbitrary element of the algebra of $\mathcal{E}(1, n)$, then irreducible Hermitian representation of $Z$ on a state $|\psi, m\rangle \in \mathsf{H}^{\varrho^-}$ is given by

$$\langle \delta_x, m' | \varrho'(Z) | \psi, m \rangle = \langle \delta_x, m' | \sigma'(Z) \oplus \rho'(Z) | \psi, m \rangle =$$
$$\langle \delta_x | \psi \rangle \langle m' | \sigma'(Z) | m \rangle \oplus \delta_{m',m} \langle \delta_x | \rho'(Z) | \psi \rangle$$

The representation of the basis is then

$$\langle \delta_x, m' | \varrho'(J_k) | \psi, m \rangle = \langle \delta_x | \psi \rangle \langle m' | \sigma'(J_k) | m \rangle = \Sigma^k_{m',m} \psi_m(q^i, t)$$
$$\langle \delta_x, m' | \varrho'(K_k) | \psi, m \rangle = 0$$
$$\langle \delta_x, m' | \varrho'(E) | \psi, m \rangle = \delta_{m',m} \langle \delta_x | \rho'(E) | \psi \rangle = \delta_{m',m} \, i \frac{\partial}{\partial t} \psi_m(q^i, t)$$
$$\langle \delta_x, m' | \varrho'(P_k) | \psi, m \rangle = \delta_{m',m} \langle \delta_x | \rho'(P_k) | \psi \rangle = -\delta_{m',m} \, i \frac{\partial}{\partial q^k} \psi_m(q^i, t)$$

## 2.3 The Poincaré field equations

The particle states in the standard 3+1 Poincaré theory are given by the states $|\psi\rangle$ in the Hilbert space that simultaneously diagonalize the unitary irreducible representations of the Casimir operator field equations.

$$\varrho'(C_\alpha)|\psi\rangle = c_\alpha |\psi\rangle, \ \alpha = 2, 4, \ c_\alpha \in \mathbb{R}$$

where

$$|\psi\rangle \in \mathsf{H}^\varrho \simeq \mathfrak{V}^{\dim \sigma} \otimes \mathcal{L}^2(\mathbb{H}_-, \mathbb{C})$$

As noted in the previous section, the representation theory of physical interest has two cases depending on whether $c_1$ is negative or zero corresponding to the timelike (massive) case and the null (massless case.)

Consider first the timelike field equations that have $c_2 < 0 \in \mathbb{R}$ and so $\mu^2 = -c_2$ may be taken to be a positive real number characterizing the mass.

$\sigma$ are the unitary irreducible representations of $\mathcal{SO}(3)$. The lift to the algebra $\sigma'$ is described above in (2.5) and so $\dim \sigma = 2j+1$.

Define the 3 notation generators

$$E = Y_0, \quad P_i = Y_i, \quad J_i = \epsilon_i^{j,k} L_{j,k}, \quad K_i = L_{0,i}$$

The Casimir invariant $W^2$ in terms of these generators is

$$W^2 = -J^2 E^2 + (P_i J^i)^2 + K^2 P^2 - (K^i P_i)^2$$

And thus the representation of the Casimir invariants are

$$\varrho'(C_2) |\psi\rangle = (-\xi'(E)^2 + (\xi'(P_i))^2) |\psi\rangle$$
$$\varrho'(C_4) |\psi\rangle = \varrho'(W^2) |\psi\rangle = \left(-\sigma'(J)^2 \xi'(E)^2 + (\xi'(P_i) \sigma'(J^i))^2\right) |\psi\rangle$$

where it has been noted that $\sigma'(K_i) |\psi\rangle = 0$.

Defining $\langle m| \sigma'(J^i) |m'\rangle = (\Sigma^i)_{m,m'}$ with $m, m' = -j, \ldots 0, \ldots j$ we have the matrix elements of the Casimir equations

$$\langle m, \delta_x | \varrho'(C_2) |\psi\rangle = c_2 \psi_m(t, q) = \left(\delta_{m,m'} \tfrac{\partial^2}{\partial t^2} - \delta_{m,m'}\left(\tfrac{\partial}{\partial q^i}\right)^2\right) \psi_{m'}(t, q)$$
$$\langle m, \delta_x | \varrho'(C_4) |\psi\rangle = c_4 \psi_m(t, q) = \left((\Sigma^2)_{m,m'} \tfrac{\partial^2}{\partial t^2} - ((\Sigma^i)_{m,m'} \tfrac{\partial}{\partial q^i})^2\right) \psi_{m'}(t, q)$$

The case $j = 0$ yields the scalar Klein Gordon equation for which $(\Sigma^i)_{m,m'} = 0$ with $m, m' = 0$. Then, $c_4 = 0$ and the first equation is simply

$$-\eta^{a,b} \tfrac{\partial^2}{\partial x^a \partial x^b} \psi(x) = \Box \psi(x) = \mu^2 \psi(x)$$

where $\Box = -\eta^{a,b} \tfrac{\partial^2}{\partial x^a \partial x^b}$.

The case $j = \tfrac{1}{2}$ yields the Dirac equation. In this case the $\Sigma$ are just the Pauli matrices and so in this case we adopt the customary notation and set $\Sigma^i = \sigma^i$ (not to be confused with the representation function $\sigma$) and suppress the $m, m'$ indices over the 2 dimensional vector space $\mathcal{V}^2$ giving

$$\left(\tfrac{\partial^2}{\partial t^2} - \left(\tfrac{\partial}{\partial q_i}\right)^2\right) \psi(t, q) = c_2 \psi(t, q)$$
$$\left(\sigma^2 \tfrac{\partial^2}{\partial t^2} - \left(\sigma^i \tfrac{\partial}{\partial q_i}\right)^2\right) \psi(t, q) = c_4 \psi(t, q)$$

Note

$$\sigma^i \sigma^j \tfrac{\partial^2}{\partial q^i \partial q^j} = \tfrac{1}{2} (\sigma^i \sigma^j + \sigma^j \sigma^i) \tfrac{\partial^2}{\partial q^i \partial q^j} = \tfrac{1}{2} \delta^{i,j} \tfrac{\partial^2}{\partial q^i \partial q^j}$$

as the anti-commutation of the $\sigma^i$ is a a property of the representations of $\mathcal{SO}(3) \simeq \mathcal{SU}(2)/\mathbb{Z}_2$.

This is solved with the eigenvalues, $c_2 = c_4 = -\mu^2$ and

$$\left(1 \tfrac{\partial}{\partial t} + \sigma^i \tfrac{\partial}{\partial q^i}\right)\psi^+(t, q) = \pm\mu\, \psi^-(t, q)$$
$$\left(1 \tfrac{\partial}{\partial t} - \sigma^i \tfrac{\partial}{\partial q^i}\right)\psi^-(t, q) = \pm\mu\, \psi^+(t, q)$$

where 1 is the unit $2 \times 2$ matrix. Now it follows the standard argument. Putting the $\psi_m^-(t, q)$, $\psi_m^+(t, q)$ with $m = 1, 2$ into a single 4 tuple $\psi_r(t, q)$ with $r, s.. = 1, ..4$ gives the usual Dirac equation

$$\left(1 \tfrac{\partial}{\partial t} + \alpha^i \tfrac{\partial}{\partial q^i}\right)\psi(t, q) = \mu\, \beta\, \psi(t, q)$$

where

$$\psi(t, q) = \begin{pmatrix} \psi^-(t, q) \\ \psi^+(t, q) \end{pmatrix}, \alpha^i = \begin{pmatrix} 0 & \sigma^i \\ -\sigma^i & 0 \end{pmatrix} \text{ and } \beta^i = \begin{pmatrix} \pm 1 & 0 \\ 0 & \pm 1 \end{pmatrix}$$

and 1 is now the unit $4 \times 4$ matrix. .

The exact same arguments may be used with the null little group $\mathcal{E}(2)$ to obtain Maxwell's equations.

## 2.4 Heisenberg group

The Heisenberg group is $\mathcal{H}(n) = \mathcal{T}(n) \otimes_s \mathcal{T}(n+1)$ with the Lie algebra generators $\{Q_i, P_i, I\}$ satisfying the Heisenberg algebra

$$[Q_i, P_j] = \delta_{i,j} I$$

where we continue to work in natural units $c = b = \hbar = 1$ and note the note about the absent $i^{[9]}$. In units where this is not the case and the generators $Q_i$ have units of length and $P_j$ have units of momentum, then

$$[Q_i, P_j] = \lambda_q \lambda_p \delta_{i,j} I = \hbar\, \delta_{i,j} I$$

Returning to natural scales, it can be directly calculated that there is a single Casimir invariant $I$ for the algebra.

It is also convenient to define the complex coherent basis $\{A_i^\pm, I\}$ of the Heisenberg algebra by

$$A_i^\pm = \tfrac{1}{\sqrt{2}} (Q_i \pm i P_i)$$

and

$$[A_i^+, A_j^-] = -i\,\delta_{i,j}\,I$$

The representations $\varrho$ of $\mathcal{H}(n)$ may be computed using the Mackey theory with an abelian normal subgroup $\mathcal{T}(n+1)$. (Note that we always use the same symbols $\varrho, \sigma, \rho, \xi$ for the various representations that appear in the Mackey theorem. The group to which they apply is should be clear from the context and so they are not distinguished further. This is also true for the Casimir invariants and their eigenvalues $C_1, c_1$)

The representation theory has two basic cases that may be categorized by whether the representation $\varrho$ of the Casimir invariant is zero or nonzero. If $\varrho'(C_1) = \varrho'(C_1) = 0$, then the representations degenerate to those of the abelian case $\mathcal{T}(2n)$ that we do not consider further. We consider only the quantum representations for which $\varrho'(I)|\psi\rangle = c_1|\psi\rangle$ with $c_1 \neq 0$.

Application of the Mackey theory in this case shows that the little group is the trivial group $e$ containing only the identity element $e$. The stabilizer is $e \otimes_s \mathcal{T}(n+1) \simeq \mathcal{T}(n+1)$.

The next step in the Mackey theory is to determine the Hilbert space on which the representation acts. The Hilbert space $\mathsf{H}^\xi$ is the $\mathcal{L}^2$ space over the symmetric space constructed as the quotient of the full group with its stabilizer which in this case is

$$\mathcal{T}(n) \otimes_s \mathcal{T}(n+1) / \mathcal{T}(n+1) \simeq \mathbb{R}^n$$

and so $\mathsf{H}^\xi = \mathcal{L}^2(\mathbb{R}^n, \mathbb{R})$. As the little group is trivial, its representations $\sigma$ is trivial and so the Hilbert space is the space with a single point. Thus, we have

$$\mathsf{H}^\varrho = \mathsf{H}^\xi = \mathcal{L}^2(\mathbb{R}^n, \mathbb{R})$$

2.4.1 Representation of the algebra

The Hermitian representation of the algebra $\varrho'$ satisfy the familiar commutation relations

$$[\varrho'(Q_i), \varrho'(P_j)] = i\,\delta_{i,j}\,\varrho'(I)$$

where $\varrho'(Q_i)$, $\varrho'(P_i)$ and $\varrho'(I)$ are Hermitian operators on $\mathsf{H}^\varrho \simeq \mathcal{L}^2(\mathbb{R}^n, \mathbb{R})$. The components with respect an orthonormal position coordinate basis $|\delta_q\rangle$ is the very familiar

$$\langle \delta_q | \varrho'(Q_i) | \psi \rangle = q^i \psi(q), \quad \langle \delta_q | \varrho'(P_i) | \psi \rangle = -i\,\frac{\partial}{\partial q^i}\psi(q) \tag{2.6}$$

Equivalently, the representation may be computed relative to a momentum coordinate basis $|\epsilon_p\rangle$ is

$$\langle \epsilon_p | \varrho'(Q_i) | \psi \rangle = -i \frac{\partial}{\partial p^i} \tilde{\psi}(p), \quad \langle \epsilon_p | \varrho'(P_i) | \psi \rangle = p^i \tilde{\psi}(p) \tag{2.7}$$

The unitary transformation between these orthonormal bases is the Fourier transform and $|\epsilon_p\rangle = \mathsf{f} |\delta_q\rangle$ and its inverse $|\delta_q\rangle = \mathsf{f}^{-1} |\epsilon_p\rangle$.

Of course any orthonormal basis could be used and in particular the coherent basis $|\eta_k\rangle$ that is its own Fourier transform is of considerable value $|\eta_k\rangle = \mathsf{f} |\eta_k\rangle$.

Now, as any orthonormal basis of the Hilbert space may be used to compute the matrix elements of the operators in Dirac's transformation theory consider another basis $|\eta_k\rangle$ of interest has the property that it is its own Fourier transform, $|\eta_k\rangle = \mathsf{f} |\eta_k\rangle$. The component functions may be computed in a $|\delta_x\rangle$ or $|\epsilon_x\rangle$ basis. First define the one dimensional case

$$\langle \delta_s | \eta_i \rangle = \eta_i(s) = \frac{1}{\sqrt{\pi \, 2^i \, i!}} e^{-\frac{1}{2} s^2} H_i(s), \; i \in \mathbb{N}, \; s \in \mathbb{R} \tag{2.8}$$

and the $n$ dimensional cases is

$$\langle \delta_q | \eta_K \rangle = \eta_K(q) = \eta_{k_1}(q^1) \dots \eta_{k_n}(q^n)$$

with $q = \{q^1, .. q^n\} \in \mathbb{R}^n$ and $K = \{k_1 ... k_n\} \in \mathbb{N}^n$. As it is its own Fourier transform, $\langle \epsilon_p | \eta_k \rangle = \eta_k(p)$ also.

In the coherent basis $\{A_i^\pm, I\}$ the representation is

$$\begin{aligned}\langle \eta_k | \varrho'(A_i^+) | \psi \rangle &= \sqrt{k_i + 1} \; \psi_{k_1 \dots k_i+1, \dots k_n}, \\ \langle \eta_k | \varrho'(A_i^-) | \psi \rangle &= \sqrt{k_i} \; \psi_{k_1 \dots k_i-1, \dots k_n}\end{aligned} \tag{2.9}$$

As there is a single Casimir invariant $I$ there are no interesting field equations to discuss. This should not be surprising as the Heisenberg group plays the role of a *translation* group on this nonabelian quantum space. Clearly it is the semidirect product of two translation groups and $Q_i$ and $P_i$ are both *translation* generators. It is just that the cannot be simultaneously diagonalized in the representation $\varrho'$ acting on $\mathcal{L}^2(\mathbb{R}^n, \mathbb{C})$.

The above results generalize directly to $\mathcal{H}(1, n)$ with the addition of the *time* generator $T$ and the *energy* generator $E$. First, the Heisenberg group has no notion of an orthogonal metric and so $\mathcal{H}(1, n) \simeq \mathcal{H}(1 + n)$. The notation is a convenience for when it is a normal subgroup of the group $\mathcal{C}(1, n)$ for which an orthogonal metric is defined. The additional generators satisfy the relation

$$[T, E] = -I$$

or in units that are not the natural units $[T, E] = -\hbar I$ as $\lambda_t \lambda_e = \hbar$ also. The representation with time diagonal is

$$\langle \delta_t | \varrho'(T) | \psi \rangle = t\psi(t), \langle \delta_t | \varrho'(E) | \psi \rangle = i\frac{\partial}{\partial t}\psi(t) \tag{2.10}$$

or energy

$$\langle \delta_t | \varrho'(T) | \psi \rangle = i\frac{\partial}{\partial e}\tilde{\psi}(e), \langle \delta_t | \varrho'(E) | \psi \rangle = e\tilde{\psi}(e)$$

This can obviously be cast in four notation with position and time coordinates to be diagonalized. Define $\{X_a\} \simeq \{T, Q_i\}$, $\{Y_a\} \simeq \{E, P_i\}$ and $\{x^a\} \simeq \{t, q^i\}$ then

$$\langle \delta_x | \varrho'(X_a) | \psi \rangle = x^a \psi(x), \langle \delta_x | \varrho'(Y_a) | \psi \rangle = -i\eta^{a,b}\frac{\partial}{\partial x^b}\psi(x)$$

where $\psi(x) = \psi(t, q)$. However, we could have equally well chosen the momentum coordinates to be diagonalized. In this case, define $\{X_a\} \simeq \{T, Q_i\}$, $\{Y_a\} \simeq \{E, -P_i\}$ and $\{x^a\} \simeq \{t, -p^i\}$ then, again

$$\langle \delta_x | \varrho'(X_a) | \psi \rangle = x^a \psi(x), \langle \delta_x | \varrho'(Y_a) | \psi \rangle = -i\eta^{a,b}\frac{\partial}{\partial x^b}\psi(x)$$

where $\psi(x) = \psi(t, p)$. This is likewise true for the energy-momentum case ($\psi(x) = \psi(e, p)$) or the energy-position ($\psi(x) = \psi(e, q)$) coordinates diagonalized. There is nothing intrinsic in the representations of the Heisenberg group that prefers which of the degrees of freedom are diagonal in the unitary irreducible representations acting on the Hilbert space $\mathsf{H}^\varrho = \mathsf{H}^\xi = \mathcal{L}^2(\mathbb{R}^{n+1}, \mathbb{C})$.

Finally, for the coherent representation we note that

$$A_0^\pm = \frac{1}{\sqrt{2}}(T \pm iE)$$

and $[A_0^+, A_0^-] = iI$

$$\langle \eta_k | \varrho'(A_0^+) | \psi \rangle = \sqrt{k_0}\,\psi_{k_0-1, k_1...k_n},$$
$$\langle \eta_k | \varrho'(A_0^-) | \psi \rangle = \sqrt{k_0+1}\,\psi_{k_0+1, k_1.....k_n} \langle \delta_q | \eta_k \rangle = \eta_k(q) = \eta_{k_1}(q^1)....\eta_{k_n}(q^n)$$

Again, $\langle \delta_{t,q} | \eta_k \rangle = \eta_k(t, q) = \eta_{k_0}(t)\eta_{k_1}(q^1)....\eta_{k_n}(q^n)$ and similarly for the other bases.

# 3: The quaplectic dynamical group

## 3.1 Introduction

The canonically relativistic theory is based on the unitary irreducible representations $\varrho$ of the dynamical group $C(1, n) = \mathcal{U}(1, 3) \otimes \mathcal{H}(1, n)$ acting on the Hilbert space $\mathsf{H}^\varrho$. Just as in the case of the Poincare group, the particle states $|\psi\rangle$ are elements of this Hilbert space defined by these representations. Again, the representations of the invariant Casimir operators define field equations and the eigenvalue equations are simultaneously solved to define the particle states.

It is expected that the dynamical symmetry described fully manifests itself in the highly interacting Planck scale regime. In this regime, the theory fully exhibits the unification of the position, time, momentum and energy nonabelian degrees of freedom under the $C(1, n)$ dynamical symmetry. Particle states $|\psi\rangle$ in this theory are for this highly interacting regime rather than the free particle regime of the Poincaré group.

We will first review the properties of the algebra in the compact *complex* basis and outline the Mackey results, taking into account the non-abelian nature of $\mathcal{H}(1, n)$ for the unitary irreducible representations in the form lifted to the algebra. Next, we examine the Casimir invariant operators and the associated scalar and simplest non-scalar field equations. This theory has 5 Casimir invariant operators for $n = 3$ and so the full set of field equations is rather rich.

## 3.2 Properties of the algebra

### 3.2.1 Complex basis of algebra and Casimir invariant operators

The algebra of $C(1, n) = \mathcal{U}(1, n) \otimes \mathcal{H}(1, n)$ in complex form is

$$
\begin{aligned}
&[Z_{a,b}, Z_{c,d}] = i(Z_{c,b}\, \eta_{a,d} - Z_{a,d}\, \eta_{b,c}), \quad [A_a^+, A_b^-] = i\, \eta_{b,c}\, I, \\
&[Z_{a,b}, A_c^\pm] = \mp i\, \eta_{a,c}\, A_b^\pm .
\end{aligned}
\tag{3.1}
$$

where as usual, the indices $a, b = 0, 1, \ldots n$. $\{Z_{a,b}\}$ span the algebra of $\mathcal{U}(1, n)$ and $\{A_c^\pm, I\}$ span the algebra of $\mathcal{H}(1, n)$.

The compact case of $C(n) = \mathcal{U}(n) \otimes \mathcal{H}(n)$ is immediately obtained by restricting the indices $a, b$ to $1, \ldots n$. This will be denoted by using the indices $i, j = 1, \ldots n$. Note that under this restriction, $\eta_{i,j} = \delta_{i,j}$. The generators $\{Z_{i,j}, A_k^\pm, I\}$ span the algebra of the subgroup $C(n)$ of $C(1, n)$. The commutation relations are as above, again noting that $\eta_{i,j} = \delta_{i,j}$. All of the expressions that follow are valid for the $C(n)$ on restricting the indices $a, b = 0, 1, \ldots n$ to the indices $i, j = 1, \ldots n$.

Define

$$U \doteq \eta^{a,b} Z_{a,b} \tag{3.2}$$

This is the $\mathcal{U}(1)$ generator in the decomposition $C(1, n) = \mathcal{SU}(1, n) \otimes_s \mathcal{O}s(1, n)$ with $\mathcal{O}s(1, n) = \mathcal{U}(1) \otimes_s \mathcal{H}(1, n)$. The corresponding generators of the algebra of $\mathcal{SU}(1, n)$ are $\{\hat{Z}_{a,b}\}$

$$Z_{a,b} = \hat{Z}_{a,b} + \frac{1}{n+1} U \eta_{a,b} \tag{3.3}$$

Note that $\eta^{a,b} \hat{Z}_{a,b} = 0$ and that the $Z_{a,b}$ and $\hat{Z}_{a,b}$ commute with $U$ and the $\hat{Z}_{a,b}$ satisfy the same commutation relations as defined above in (3.1).

The Casimir invariant operators $C_\alpha$ are elements of the universal enveloping algebra $\mathfrak{e}(\mathcal{G})$ of the group $\mathcal{G}$. The number of independent invariants is equal to the rank of the group. The rank of $C(1, n)$ is $n + 2$ for $n \leq 3$[4]. Conjecturing this is true for general $n$, the $n + 2$ Casimir invariant operators are [40]

$$\begin{aligned} C_1 &= I \\ C_2 &= W = \eta^{a_1,a_2} W_{a_1,a_2} \\ C_4 &= W_2 = \eta^{a_1,a_4} \eta^{a_2,a_3} W_{a_1,a_2} W_{a_3,a_4} \\ &\ldots \\ C_{2n} &= W_n = \eta^{a_1,a_{2n}} \ldots \eta^{a_{2n-2},a_{2n-1}} W_{a_1,a_2} \ldots W_{a_{2n-1},a_{2n}} \end{aligned} \tag{3.4}$$

with

$$W_{a,b} \doteq A^+_a A^-_b - I Z_{a,b}$$

Note in particular that the second order invariant is of the form

$$C_2 = W \doteq A^2 - I U \tag{3.5}$$

with $A^2 \doteq \eta^{a,b} A^+_a A^-_b$ and $U$ is the generator of the algebra of $\mathcal{U}(1)$ defined in (3.2)

The commutation relations for the $W_{a,b}$ are

$$[Z_{a,b}, W_{c,d}] = i(\eta_{a,d} W_{b,c} - \eta_{b,c} W_{d,a}), \quad [A_c^\pm, W_{a,b}] = 0.$$

and therefore $W_{c,d}$ are Heisenberg *translation* invariant. It is important to note that both of the terms in $W_{a,b}$ are required in order for the commutator to vanish with $A_c^\pm$. The $W_{c,d}$ commutators with $Z_{a,b}$ are the same as $Z_{c,d}$. The Casimir invariants of $\mathcal{U}(1, n)$ are

$$D_1 = U = \eta^{a_1,a_2} Z_{a_1,a_2}$$
$$D_2 = Z^2 = \eta^{a_1,a_4} \eta^{a_2,a_3} Z_{a_1,a_2} Z_{a_3,a_4}$$
...
$$D_n = Z^n = \eta^{a_1,a_{2n}} ... \eta^{a_{n-1},a_n} Z_{a_1,a_2} ... Z_{a_{2n-1},a_{2n}}$$

Therefore, (3.4) are invariant unitary $\mathcal{U}(1, n)$ rotations and, as the $W_{a,b}$ have already been established to be Heisenberg translational invariant, it follows that they are Casimir invariants of $\mathcal{C}(1, n)$.

As an aside that will be useful, we note also that $\mathcal{O}s(1, n)$ is rank 2 for all $n$. The two Casimir invariants $C°_1$ and $C°_2$ (the ° distinguishing them from the Casimir invariants of $\mathcal{C}(1, n)$) are

$$C°_1 = C_1 = I$$
$$C°_2 = C_2 = A^2 - I U$$

Higher order invariants are not defined for the oscillator group as, noting (3.3) with $\hat{Z}_{a,b} = 0$, that $W°_{a,b} = A^+{}_a A^-{}_b - I U / n$ is not Heisenberg *translation* invariant.

The form of the Casimir invariants given in (3.4), while easy to write out in a generic form, do not lend themselves readily to computations. The lie algebra bracket relations (3.1) may be used to write these in a more convenient form

$$(A^-)_a \cdot Z_{a,b} \cdot (A^+)_b =$$
$$(\tilde{C}_4 + (-1 - (\tfrac{1}{2} n + 1) I + C_2) I U + \tfrac{1}{2} I^2 (1 + \tfrac{1}{n+1}) U^2) + \tfrac{1}{2} Z^2$$

with

$$\tilde{C}_4 = \tfrac{1}{2} (2n - C_2(n-1)) I - 2 C_2 + C_2^2 - C_4$$

$U$ and $Z^2$ are the Casimir invariants of the unitary group, $D_1 = U$ and $D_2 = Z^2$. Noting that $I = C_1$, this may be written in a more general form as

$$\eta^{a,b} \eta^{c,d} A_a^- Z_{b,c} A_d^+ = \sum_{\alpha=1}^{2} \sum_{\kappa=0}^{2-\alpha+1} f_{\kappa,\alpha}^2(n, C_1, C_2) D_\alpha{}^\kappa \qquad (3.7)$$

with the coefficient functions depending only on the Casimir invariants of the quaplectic group

$$f_{0,1}^2(n, C_1, C_2) = \tfrac{1}{2} (2n - C_2(n-1) C_1 - 2 C_2 + C_2^2 - C_4)$$
$$f_{1,1}^2(n, C_1, C_2) = (-1 - (\tfrac{n}{2} + 1) C_1 + C_2)$$
$$f_{2,1}^2(n, C_1, C_2) = \tfrac{1}{2} C_1^2 (1 + \tfrac{1}{n+1}) \qquad (3.8)$$
$$f_{1,2}^2(n, C_1, C_2) = \tfrac{1}{2}$$

Note that this equation is a second order equation in the operators $\{A_a^\pm\}$. Similar expressions may be obtained for the remaining $C_\alpha$ and it may be verified that, at least up to $C_8$, these equations are also second order equations in terms of the operators $\{A_a^\pm\}$. In fact, the equation for $C_{2m}$ with $m = 1, ..n$ is given by

$$\eta^{a_1,a_{2m}} ... \eta^{a_{2m-2},a_{2m-1}} A_{a_1}^- Z_{a_2,a_3} .. Z_{a_{2m-2},a_{2m-1}} A_{a_{2m}}^+ = \\ \sum_{\alpha=1}^{m} \sum_{\kappa=0}^{m-\alpha+1} f_{\kappa,\alpha}^m(n+1, C_1, C_2, ... C_{2m}) D_\alpha^{\;\kappa} \tag{3.9}$$

It may be conjectured that for the group $C(1, n)$ that expressions of this form hold for $m = 1, ... n + 1$ but not this has only been established by direct computation for $n \leq 3$.

3.2.2   Real $n + 1$ basis for the quaplectic algebra

Define

$$A_a^\pm = \tfrac{1}{\sqrt{2}}(X_a \mp i\, Y_a), \quad Z_{a,b} = \tfrac{1}{2}(M_{a,b} - i\, L_{a,b})$$

A general element of the quaplectic algebra is $Z + A$ with $Z$ and element of the algebra of $\mathcal{U}(1, n)$ and $A$ an element of the Heisenberg algebra

$$Z = \phi^{a,b} M_{a,b} + \varphi^{a,b} L_{a,b} \\ A = \iota I + x^a X_a + y^a Y_a$$

These generators satisfy

$$[L_{a,b}, L_{c,d}] = -L_{b,d}\,\eta_{a,c} + L_{b,c}\,\eta_{a,d} + L_{a,d}\,\eta_{b,c} - L_{a,c}\,\eta_{b,d}, \\ [L_{a,b}, M_{c,d}] = -M_{b,d}\,\eta_{a,c} - M_{b,c}\,\eta_{a,d} + M_{a,d}\,\eta_{b,c} + M_{a,c}\,\eta_{b,d} \\ [M_{a,b}, M_{c,d}] = -L_{b,d}\,\eta_{a,c} - L_{b,c}\,\eta_{a,d} - L_{a,d}\,\eta_{b,c} - L_{a,c}\,\eta_{b,d} \\ \\ [L_{a,b}, X_c] = -X_b\,\eta_{a,c} + X_a\,\eta_{b,c}, \quad [L_{a,b}, Y_c] = -Y_b\,\eta_{a,c} + Y_a\,\eta_{b,c}, \\ [M_{a,b}, X_c] = -Y_b\,\eta_{a,c} - Y_a\,\eta_{b,c}, \quad [M_{a,b}, Y_c] = X_b\,\eta_{a,c} + X_a\,\eta_{b,c} \\ [X_a, Y_b] = I\,\eta_{a,c} \tag{3.10}$$

Clearly both the generators $\{L_{a,b}, X_c\}$ and $\{L_{a,b}, Y_c\}$ define the algebra for the Poincaré subgroups.

For notational convenience, define $X^2 = \eta^{a,b} X_a X_b$ and $Y^2 = \eta^{a,b} Y_a Y_b$. The second order Casimir Invariant is

$$C_2 = (X^2 + Y^2 + I - I\,U)$$

The term $I$ is the first order Casimir invariant $C_1 = I$ and as any linear combination of Casimir invariants is a Casimir invariant, this term may just be dropped. We will always do this in what follows.

$$C_2 = \tfrac{1}{2}(X^2 + Y^2) - I\,U$$

In the introduction, the limiting form of the quaplectic algebra for small velocities and interactions was described $b, c \to \infty$. This is important to make connection with existing theory. The contraction of the representations, although computationally more complex, must follow the contraction of the abstract algebra.

It is useful to consider only the limit $b \to \infty$ corresponding to small interactions. One would expect this to contract essentially to the Poincaré group. Choose units in which $c = \hbar = 1$ but $b$ is explicit. In these units, $\lambda_t = \lambda_q = \frac{1}{\sqrt{b}}$ and $\lambda_e = \lambda_p = \sqrt{b}$. Again, as in (1.6), the limit $b \to 0$ is given by $\phi^{a,b} \to r^{a,b}/b$ and $M_{a,b} \to b M^\circ{}_{a,b}$. Substituting into the algebra (3.10) gives the nonzero generators for the contracted algebra in the limit $b \to \infty$

$$[L_{a,b}, L_{c,d}] = -L_{b,d}\,\eta_{a,c} + L_{b,c}\,\eta_{a,d} + L_{a,d}\,\eta_{b,c} - L_{a,c}\,\eta_{b,d}$$
$$[L_{a,b}, M^\circ{}_{c,d}] = -M^\circ{}_{b,d}\,\eta_{a,c} - M^\circ{}_{b,c}\,\eta_{a,d} + M^\circ{}_{a,d}\,\eta_{b,c} + M^\circ{}_{a,c}\,\eta_{b,d}$$
$$[M^\circ{}_{a,b}, Y_c] = X_b\,\eta_{a,c} + X_a\,\eta_{b,c} \quad (3.11)$$
$$[L_{a,b}, X_c] = -X_b\,\eta_{a,c} + X_a\,\eta_{b,c},\ [L_{a,b}, Y_c] = -Y_b\,\eta_{a,c} + Y_a\,\eta_{b,c}$$
$$[X_a, Y_b] = I\,\eta_{a,b}$$

The generators $M^\circ{}_{a,b}$ now generate an abelian subgroup $\mathcal{M}(n+1)$ where $\mathcal{M}(n) \simeq \mathbb{R}^N$ with $N = n(n+1)/2$. Inspection of the algebra shows that it is the algebra of the group

$$\lim_{b \to \infty} C^\hbar(1, n) = \mathcal{B}(1, n) \simeq$$
$$\mathcal{SO}(1, n) \otimes_s \mathcal{M}(n+1) \otimes_s \mathcal{H}(n+1) \simeq \mathcal{E}(1, n) \otimes_s \mathcal{M}(n+1) \otimes_s \mathcal{T}(n+1)$$

where in the second form, the generators of $\mathcal{E}(1, n)$ are $\{L_{a,b}, Y_b\}$ and $\mathcal{T}(n+1)$ are $\{X_a, I\}$. The number of Casimir invariants of $\mathcal{B}(1, n)$ can be verified by direct computation to be $n + 2$ for $n \leq 3$[4]. Two of these invariants are

$$C_1 = I,\ C_2 = \tfrac{1}{2} X^2 - I U$$

The form of the remaining invariants remains to be determined.

3.2.3 Real $n$ basis for the quaplectic algebra

This section considers writes the algebra in an $n$ notation. For $n = 3$, this is of course a *three* notation as apposed to the above *four* notation. This formulation, while more explicit, has the value that it makes the canonical *quad symmetry*, particular the duality between the $\{T, Q_i\}$ and $\{T, P_i\}$ or $\{E, P_i\}$ and $\{E, Q_i\}$ more explicit. Define

$$K_i = -L_{0,i},\ N_i = M_{0,i},\ J_i = \epsilon_i^{j,k} L_{j,k},\ R = \tfrac{1}{2} M_{0,0}$$
$$X_i = Q_i,\ Y_i = P_i,\ X_0 = T,\ Y_0 = E,$$

and $M_{i,j}$ remains itself. As always, $i, j = 1, ..n$. Then, with dimensional scales explicit, a general element of the quaplectic algebra is $Z + A$ where

$$Z = \vartheta R + \beta^i K_i + \gamma^i N_i + \theta^{i,j} M_{i,j}$$

$$A = \iota I + \frac{1}{\lambda_t} t T + \frac{1}{\lambda_e} e E + \frac{1}{\lambda_q} q^i Q_i + \frac{1}{\lambda_p} p^i P_i$$

The corresponding commutator relations for the algebra of the subgroup $C(n)$ are given directly by (3.10)

$$[J_{i,j}, J_{k,l}] = -J_{j,l}\delta_{i,k} + J_{j,k}\delta_{i,l} + J_{i,l}\delta_{j,k} - J_{i,k}\delta_{j,l},$$
$$[J_{i,j}, M_{k,l}] = -M_{j,l}\delta_{i,k} - M_{j,k}\delta_{i,l} + M_{i,l}\delta_{j,k} + M_{i,k}\delta_{j,l}$$
$$[M_{i,j}, M_{k,l}] = -J_{j,l}\delta_{i,k} - J_{j,k}\delta_{i,l} - J_{i,l}\delta_{j,k} - J_{i,k}\delta_{j,l}$$

$$[J_{i,j}, Q_k] = -Q_j\delta_{i,k} + Q_i\delta_{j,k}, \quad [J_{i,j}, P_k] = -P_j\delta_{i,k} + P_i\delta_{j,k},$$
$$[M_{i,j}, Q_k] = -\frac{c}{b}(P_j\delta_{i,k} + P_i\delta_{j,k}), \quad [M_{i,j}, P_c] = \frac{b}{c}(Q_j\delta_{i,k} + Q_i\delta_{j,k})$$
$$[Q_i, P_j] = I\hbar\delta_{i,j}$$

(3.12)

The remaining commutator relations are

$$[K_i, K_k] = J_{i,k}, \quad [J_{i,j}, K_k] = -K_j\delta_{i,k} + K_i\delta_{j,k},$$
$$[N_i, N_k] = J_{i,k}, \quad [J_{i,j}, N_k] = -N_j\delta_{i,k} + N_i\delta_{j,k},$$

$$[K_i, R] = N_i, \quad [K_i, M_{k,l}] = N_l\delta_{i,k} N_k\delta_{i,l}, \quad [K_i, P_j] = -E\delta_{i,j},$$
$$[N_i, R] = K_i, \quad [N_i, M_{k,l}] = K_l\eta_{i,k} + K_k\eta_{i,l}, \quad [N_i, P_j] = T\delta_{i,j}$$

$$[K_i, T] = \frac{1}{c}Q_i, \quad [K_i, E] = cP_i, \quad [K_i, Q_j] = \frac{1}{c}T\delta_{i,j},$$
$$[N_i, T] = \frac{1}{b}P_i, \quad [N_i, E] = -bQ_i, \quad [N_i, Q_j] = -\frac{1}{b}E\delta_{i,j},$$

(3.13)

$$[K_i, P_j] = \frac{1}{c}E\delta_{i,j}, \quad [K_i, R] = N_i, \quad [R, T] = \frac{1}{bc}E,$$
$$[N_i, P_j] = \frac{1}{b}T\delta_{i,j}, \quad [N_i, R] = K_i, \quad [R, E] = -bcT,$$

$$[M_{i,j}, Q_k] = \frac{c}{b}(P_j\delta_{i,k} + P_i\delta_{j,k}), \quad [M_{i,j}, P_k] = -\frac{b}{c}(Q_j\delta_{i,k} + Q_i\delta_{j,k})$$

$$[K_i, N_j] = M_{i,j} + \delta_{i,j}R, \quad [T, E] = -\hbar I$$

While these equations are less compact than the complex or real four form, they are useful in performing actual computations. It also makes manifest the symmetry between the velocity boost generators $K_i$ and the momentum boost generators $N_i$ that is central to the orthogonal reciprocity described in the introduction. Substituting the scaled generators $G_i$, $F_i$ and $M°_{i,j}$ defined in (1.6) and (1.18) gives the limiting form of the algebra given in (1.18).

And again the second order Casimir invariant is (1.12)

$$C_2 = \frac{-1}{2\lambda_t^2}\left(T^2 + \frac{1}{c^2b^2}E^2 - \frac{1}{c^2}Q^2 - \frac{1}{b^2}P^2\right) - IU$$

## 3.3 Overview of the Mackey theory for the quaplectic group

The unitary irreducible representations of the quaplectic group $C(1, n)$ may be constructed using the Mackey representation theory from projective extension to $C(1, n)$ of the unitary irreducible representations of the Heisenberg group $\mathcal{H}(1, n)$ and the unitary representations of the unitary groups $\mathcal{U}(1, n)$. These representations follow exactly the same formalism as the Poincaré group but due to specific properties of the group, the application of the theory is quite different. First, the normal subgroup $\mathcal{H}(1, n)$ is nonabelian and so the extension of $\rho$ is not trivial. The stabilizer, for the quantum case where the representations of the first Casimir invariant, $\varrho'(I)$, are non trivial, the stabilizer is the group $C(1, n)$ itself and so the application of the induction theorem is trivial. On the other hand, the little group in this case is the noncompact $\mathcal{U}(1, n)$ which does not have finite dimensional unitary representations.

With these introductory remarks, we now review more systematically the unitary irreducible representations of $C(1, n)$. As noted, the representations have two general classes depending on the eigenvalue of the representation of the first Casimir invariant $I$:

$$\varrho'(I) |\psi\rangle = c_1 |\psi\rangle \text{ with } |\psi\rangle \in \mathsf{H}^\varrho$$

Before continuing, it is necessary to make a remark about notation. We have a limited number of symbols at our disposal and so continue to use the symbol $\varrho$ to designate general unitary irreducible representations of the group in question, which in this context is the quaplectic group $C(1, n)$. This is the same symbol that was used in the Poincaré case and so, strictly speaking, they should be labelled with the group in question to be explicit, $(\varrho')_{C(1,n)}(I)$, $(\varrho')_{\mathcal{E}(1,n)}(E)$ and so forth. This is clearly too cumbersome and so this is left implicit and made clear by the context. The same is true of the various other symbols in the representation theory, they are all implicitly labelled by the group in question.

### 3.3.1 Hilbert Spaces of the degenerate abelian representations

Returning to the Casimir invariant, if $c_1 = 0$, the representation reduces to the degenerate case where the representations of $C(1, n)$ degenerate to those of the complex Euclidean group $\mathcal{E}_c(1, n) = \mathcal{U}(1, n) \otimes_s \mathcal{T}_c(1, n)$ with the abelian normal complex translation subgroup $\mathcal{T}_c(1, n)$. (The $c$ subscript denotes that these are the Euclidean group with a Hermitian metric and translations on a complex vector space.) This representation is *classical* in the sense that the momentum and position generators commute in this degenerate representation. The second order Casimir invariant reduces to the simple Hermitian metric

$$\varrho'(W) |\psi\rangle = (\varrho'(A^2) + \varrho'(I)\varrho'(U)) |\psi\rangle = \varrho'(A^2) |\psi\rangle = c°_2 |\psi\rangle$$

This has the corresponding Mackey representation theory. The representations break into the cases depending on whether $\kappa°_2$ is less than, equal to or greater than zero. The corresponding little groups, stabilizer groups and symmetric spaces are

| | Little Group | Stabilizer Group | Symmetric Space |
|---|---|---|---|
| $c°_2 < 0$ | $\mathcal{U}(n)$ | $\mathcal{E}_c(n)$ | $\mathbb{P}^{n+1}_- \simeq \mathcal{E}_c(1, n)/\mathcal{E}(n) \simeq \mathcal{U}(1, n)/\mathcal{U}$ |
| $c°_2 = 0$ | $\mathcal{E}_c(n-1)$ | $\mathcal{G}_c(n)$ | $\mathbb{P}^{n+1}_0 \simeq \mathcal{E}_c(1, n)/\mathcal{G}_c(n) \simeq \mathcal{U}(1, n)/\mathcal{E}_c($ |
| $c°_2 > 0$ | $\mathcal{U}(1, n-1)$ | $\mathcal{E}_c(1, n-1)$ | $\mathbb{P}^{n+1}_+ \simeq \mathcal{E}_c(1, n)/\mathcal{E}_c(1, n-1) \simeq \mathcal{U}(1, n)/\mathcal{U}$ |

where we define the complex Galelei group as $\mathcal{G}_c(n) = \mathcal{E}_c(n-1) \otimes_s \mathcal{T}_c(1, n)$. The Hilbert spaces of the representations are defined over these complex projective spaces $\mathsf{H}^{\varrho\pm,0} \simeq \mathfrak{V}^{\dim \sigma} \otimes \mathcal{L}^2(\mathbb{P}_{\pm,0}, \mathbb{R})$ where $\mathfrak{V}^{\dim \sigma}$ is the finite or countably finite dimensional vector space for the unitary irreducible representations of the little group. Note that, like the Poincare group the generalized timelike *little* groups are compact with finite dimensional unitary irreducible representations.

Note that for the compact case $C(n)$, the degenerative case is $\mathcal{E}_c(n) = \mathcal{U}(n) \otimes_s \mathcal{T}_c(n)$ for which there is only one (non-trivial) case and the symmetric space is the complex projective spheres $\mathbb{S}_c^n \simeq \mathcal{U}(n)/\mathcal{U}(n-1) \simeq \mathbb{S}^{2n-1}$. The Hilbert space is $\mathsf{H}^\varrho \simeq \mathfrak{V}^{\dim \sigma} \otimes \mathcal{L}^2(\mathbb{S}^{2n-1}, \mathbb{R})$ where $\dim \sigma$ is the dimension of the unitary irreducible representations of the little group $\mathcal{U}(n-1)$

These representations in this degenerate case on Hilbert spaces over the projective spaces is analogous to the Poincaré theory. Again, the projective surfaces foliate $\mathbb{R}^{2n}$.

### 3.3.2 Hilbert space of the quantum representations

The general nonabelian case corresponds to $c_1 \neq 0$. In this case, the stabilizer group is simply the quaplectic group itself, $\mathcal{G}^\circ = C(1, n)$. An immediate consequence is that the Mackey induction [28] is trivial and the representations are given directly by the direct product

$$\varrho(g(\Upsilon, \omega, \iota)) |\psi\rangle = \sigma(\Upsilon) |\psi\rangle^\sigma \otimes \rho(g(\Upsilon, \omega, \iota)) |\psi\rangle^\xi \quad \text{with } \mathsf{H}^\varrho = \mathsf{H}^\sigma \otimes \mathsf{H}^\xi,$$

and $|\psi\rangle \in \mathsf{H}^\varrho$, $|\psi\rangle^\sigma \in \mathsf{H}^\sigma$, $|\psi\rangle^\xi \in \mathsf{H}^\xi$, $g(\Upsilon, \omega, \iota) \in C(1, n)$, $\Upsilon \in \mathcal{U}(1, n)$

Consider first the representations of the subgroup $C(n)$. The Hilbert space of the unitary irreducible representations $\xi$ of the Heisenberg group $\mathcal{H}(n)$ is $\mathsf{H}^\xi = \mathcal{L}^2(\mathbb{R}^n, \mathbb{R})$. The unitary irreducible representations $\sigma$ of the compact unitary group $\mathcal{U}(n)$ are finite dimensional with dimension $N = \dim \sigma$. Therefore $\mathsf{H}^\sigma \simeq \mathfrak{V}^N$ where $\mathfrak{V}^N$ is a finite dimensional vector space of dimension $\dim \sigma$. Putting it together, the Hilbert space $\mathsf{H}^\varrho$ of the representations of $C(n)$ are

$$\mathsf{H}^\varrho \simeq \mathfrak{V}^N \otimes \mathcal{L}^2(\mathbb{R}^n, \mathbb{R}), \quad N = \dim \sigma$$

The unitary irreducible representations of the group $C(1, n)$ follow similarly. The Hilbert space of the unitary irreducible representations of the Heisenberg group $\mathcal{H}(1, n)$ is $\mathsf{H}^\xi = \mathcal{L}^2(\mathbb{R}^{n+1}, \mathbb{R})$. However, in this case the unitary irreducible representations of $\mathcal{U}(1, n)$ are not finite dimensional. However, they may be decomposed into a countably infinite sum of representations of the maximally compact subgroup $\mathcal{U}(n)$. Therefore

$$\mathsf{H}^\sigma \simeq \left( \bigoplus_l^\infty \mathfrak{V}^{N_l} \right)$$

where $N_l = \dim \sigma_l$ is the dimension of the $l$th unitary irreducible representation of $\mathcal{U}(n)$. Note that representations may appear with multiplicity. We will refer to this as a *ladder* of representations in what follows. Each finite dimensional representation is a *rung* on the *ladder*.

Thus, the Hilbert space of the unitary irreducible representations of $C(1, n)$ are

$$\mathsf{H}^\varrho \simeq \bigoplus_l^\infty (\mathfrak{V}^{N_l} \otimes \mathcal{L}^2(\mathbb{R}^{n+1}, \mathbb{R})), \quad N_l = \dim \sigma_l$$

### 3.3.3 General form of the representation of the algebras

We are now able to write down the representations of the quaplectic algebra that are given by the expression for the group (3.13) lifted to the algebra. Let $Z \in \mathfrak{a}(\mathcal{U}(1, n))$, $A \in \mathfrak{a}(\mathcal{H}(1, n))$ with $V = Z + A$ and $V \in \mathfrak{a}(C(1, n))$

$$\varrho'(V)|\psi\rangle = (\sigma'(Z) + \rho'(V))|\psi\rangle = \sigma'(Z)|\psi\rangle^\sigma \otimes |\psi\rangle^\xi \oplus |\psi\rangle^\sigma \otimes \rho'(V)|\psi\rangle^\xi$$

where $|\psi\rangle^\sigma \in \mathsf{H}^\sigma \simeq \mathfrak{V}^{\dim \sigma}$, $|\psi\rangle^\xi \in \mathsf{H}^\xi \simeq \mathcal{L}^2(\mathbb{R}^{n+1}, \mathbb{R})$ and $|\psi\rangle \in \mathsf{H}^\varrho = \mathsf{H}^\sigma \otimes \mathsf{H}^\xi$. Note that $\sigma'(A) = 0$. Again, in general for the noncompact group, $\dim \sigma$ may be countably infinite.

In terms of a basis $\{Z_{a,b}, A_a^\pm, I\}$ we therefore have

$$\begin{aligned}
\varrho'(I)|\psi\rangle &= |\psi\rangle^\sigma \otimes \rho'(I)|\psi\rangle^\xi \\
\varrho'(A_a^\pm)|\psi\rangle &= |\psi\rangle^\sigma \otimes \rho'(A_a^\pm)|\psi\rangle^\xi \\
\varrho'(Z_{a,b})|\psi\rangle &= \sigma'(Z_{a,b})|\psi\rangle^\sigma \otimes |\psi\rangle^\xi \oplus |\psi\rangle^\sigma \otimes \rho'(Z_{a,b})|\psi\rangle^\xi
\end{aligned} \qquad (3.14)$$

Note the key difference, as mentioned above, that the representations $\rho'(Z_{a,b})$ are not trivial as the normal subgroup $\mathcal{H}(1, n)$ is nonabelian. In this sense, the generators $Z_{a,b}$ of the little group $\mathcal{SU}(1, n)$ appear twice, once in the representation $\sigma'$ acting on the Hilbert space $\mathsf{H}^\sigma$ that is a countably infinite vector space and $\rho'$ acting on the function space $\mathsf{H}^\xi \simeq \mathcal{L}^2(\mathbb{R}^{n+1}, \mathbb{R})$.

It remains then to compute $\rho'(I)|\psi\rangle^\xi$, $\rho'(A_a^\pm)|\psi\rangle^\xi$, $\sigma'(Z_{a,b})|\psi\rangle^\sigma$ and $\rho'(Z_{a,b})|\psi\rangle^\xi$ to determine the full representation of the algebra. This is addressed in the following sections.

## 4: The field equations of the quaplectic group $C(n)$

It is useful to first consider the representations corresponding to the $C(n)$ subgroup of $C(1, n)$ to establish the method and then to generalize to the case $C(1, n)$. The reason for this is that the little group for $C(n)$ is $\mathcal{U}(n)$ with finite dimensional unitary irreducible representations and therefore the Hilbert spaces $\mathsf{H}^\sigma \simeq \mathfrak{V}^N$ are just finite dimensional vector spaces with dimension $N = \dim \sigma$.

The steps are to first determine the unitary irreducible representations $\rho'$ of the generators $\{I, A_i^\pm, Z_{i,j}\}$ of $C(n)$ on $\mathsf{H}^\xi \simeq \mathcal{L}^2(\mathbb{R}^n, \mathbb{R})$ that is the Hilbert space for the representations of the normal subgroup $\mathcal{H}(n)$. Next the standard finite dimensional representations $\sigma'$ of $\mathcal{U}(n)$ on the finite dimensional vector space $\mathsf{H}^\sigma \simeq \mathfrak{V}^N$ are summarized. Finally, these are combined to give the full representation $\varrho'$ of the generators $\{I, A_i^\pm, Z_{i,j}\}$ of $C(n)$ on $\mathsf{H}^\varrho \simeq \mathfrak{V}^N \otimes \mathcal{L}^2(\mathbb{R}^n, \mathbb{R})$. These may be used to directly given the representations $\varrho'$ of the generators $W_{i,j} \doteq A^+{}_i A^-{}_j - I Z_{i,j}$. Noting $N_c = n + 1$, these generators may be used to construct the Casimir invariants $C_1 = I$, $C_2 = W$, $C_4 = W^2$, .. $C_{2n} = W^n$ using (3.4). The field equations are then the $N_c = n + 1$ equations

$$\varrho'(C_\alpha) |\psi\rangle = c_\alpha |\psi\rangle \quad \alpha = 1, 2, 4, .. 2n$$

with $|\psi\rangle \in \mathsf{H}^\varrho$ .

### 4.1 The representation $\rho'$ of the generators of $C(n)$.

The representations $\rho'$ of the algebra of the full group $C(n)$ act on the Hilbert space $\mathsf{H}^\xi \simeq \mathcal{L}^2(\mathbb{R}^n, \mathbb{R})$ of the representations $\xi$ of the normal subgroup $\mathcal{H}(n)$. The representations $\rho$ that are extensions of $\xi$ that appear are a direct consequence of the fact that the normal subgroup $\mathcal{H}(n)$ is nonabelian. These extensions are trivial as the Poincaré subgroup that has an abelian translation group for the normal subgroup.

The Hermitian representation $\rho'$ of the generators $\{Z_{i,j}, A_i^\pm, I\}$ of the Heisenberg subgroup $\mathcal{H}(n)$ are given by $\{Z_{i,j}, A_i^\pm, I\}$

$$\rho'(A_i^\pm) = \xi'(A_i^\pm), \quad \rho'(I) = \xi'(I)$$

These satisfy the relations given in (3.1) for a Hermitian projective representation of the quaplectic algebra

$$[\rho'(Z_{i,j}), \rho'(Z_{k,l})] = s\,(\delta_{j,k}\,\rho'(Z_{i,l}) - \delta_{i,l}\,\rho'(Z_{k,j})),$$
$$[\rho'(Z_{i,j}), \xi'(A_j^\pm)] = \pm s\,\delta_{i,k}\,\xi'(A_j^\pm) \qquad (4.1)$$
$$[\xi'(A_i^+), \xi'(A_j^-)] = -\delta_{i,j}\,\xi'(I).$$

where the appropriate factor of $i$ has been inserted into the expression for the commutation relations as the representation is Hermitian as given in (2.1). $s$ is the projective representation scaling constant.

As the $\rho'(Z_{i,j})$ are represented on the Hilbert spaces $\mathsf{H}^\xi$, these must be elements in the enveloping algebra $\mathfrak{e}(C(n))$ and so be of the form given in (2.2)

$$\varrho'(Z_{i,j}) = d_{i,j}^{k,\pm}\,\xi'(A_k^\pm) + d_{i,j}^{k,\pm,l,\pm}\,\xi'(A_k^\pm)\,\xi'(A_l^\pm) + \ldots \qquad (4.2)$$

Substituting into (4.1) gives the solution that all the $d_{i,j}^{k,\pm}, d_{i,j}^{k,\pm,l,\pm}, \ldots$ are zero except

$$d_{i,j}^{k,+,l,-} = \tfrac{1}{s}\,\delta_i^k\,\delta_j^l$$

That is,

$$\rho'(Z_{i,j}) = \tfrac{-1}{s}\,\xi'(A_i^+)\,\xi'(A_j^-) \qquad (4.3)$$

This is a specific representation of $\mathcal{U}(n)$ that is not a general representation. This can be seen by noting that the representation of the first Casimir invariant is

$$D_1 = s\,\delta^{i,j}\,\xi'(A_i^+)\,\xi'(A_j^-)$$

then the higher order $\mathcal{U}(n)$ Casimir invariants (3.6) are not independent and are just given by

$$D_\alpha = s^\alpha\,D_1(D_1 + n\,I)^{\alpha-1},\ \alpha = 2,\ ..n.$$

A basis for the Hilbert space $\mathsf{H}^\xi$ is given in (2.9) as $|\eta_\mathrm{K}\rangle = |\eta_{k_1,\ldots k_n}\rangle$, $\mathrm{K} = (k_1, \ldots k_n)$.

$$\xi'(A_i^+)\,|\eta_{k_1,\ldots k_n}\rangle = \sqrt{k_i + 1}\,|\eta_{k_1,\ldots k_i+1,\ldots k_n}\rangle$$
$$\xi'(A_i^-)\,|\eta_{k_1,\ldots k_n}\rangle = \sqrt{k_i}\,|\eta_{k_1,\ldots k_i-1,\ldots k_n}\rangle$$

Defining $\mathrm{I}_i$ to be an $n$ tuple of 0's with a 1 at the $i$th position, $\mathrm{I}_i = (0, 0, , , 1, \ldots 0)$, this may be written more compactly as

$$\xi'(A_i^+)\,|\eta_\mathrm{K}\rangle = \sqrt{k_a + 1}\,|\eta_{\mathrm{K}+\mathrm{I}_i}\rangle$$
$$\xi'(A_i^-)\,|\eta_\mathrm{K}\rangle = \sqrt{k_a}\,|\eta_{\mathrm{K}-\mathrm{I}_i}\rangle$$

It follows directly that

$$\rho'(Z_{j,i})\,|\eta_\mathrm{K}\rangle = \tfrac{1}{s}\sqrt{(k_j+1)\,k_i}\,|\eta_{\mathrm{K}-\mathrm{I}_i+\mathrm{I}_j}\rangle,\ i \ne j$$
$$\rho'(Z_{i,i})\,|\eta_\mathrm{K}\rangle = \tfrac{1}{s}\,k_i\,|\eta_\mathrm{K}\rangle,\ i = j$$

It may be verified that these satisfy the commutation relations for the projective representation of the quaplectic algebra as Hermitian operators on the Hilbert space $\mathsf{H}^\xi$ and so are, indeed, representations of the algebra.

The group may be factored as $\mathcal{U}(n) = \mathcal{U}(1) \otimes \mathcal{SU}(n)$. Note that the generator $U$ of the algebra of $\mathcal{U}(1)$ is defined by $U = \delta^{i,j} Z_{i,j}$ and it follows immediately that the representation is

$$\rho'(U) |\eta_K\rangle = (k_1 + k_2 + \ldots k_n) |\eta_K\rangle = l |\eta_K\rangle \tag{4.4}$$

with $l = \sum_{i=1}^{n} k_i$. This enables the Hilbert space $\mathsf{H}^\xi$ to be decomposed into an infinite ladder of $\mathcal{U}(n)$ invariant finite dimensional subspaces $\mathsf{H}^\xi_l$ such that

$$\mathsf{H}^\xi = \oplus_{l=1}^{\infty} \mathsf{H}^\xi_l \simeq \oplus_{l=1}^{\infty} \mathfrak{V}^l$$

The representations $\rho'(Z_{a,b})$ act on the subspaces

$$\rho'(Z_{i,j}) : \mathsf{H}^\xi_l \to \mathsf{H}^\xi_l : |\eta_K\rangle \mapsto \rho'(Z_{i,j}) |\eta_K\rangle$$

This means that the representations $\rho'$ restricted to the subspace $\mathsf{H}^\xi_l$ are the finite dimensional representations $\rho'_l$,

$$\rho'_l(Z_{i,j}) : \mathfrak{V}^l \to \mathfrak{V}^l : |\eta_K\rangle \mapsto \rho'_l(Z_{i,j}) |\eta_K\rangle$$

On the other hand the representations of the generators $\rho'(A_i^\pm)$ of $\mathcal{H}(n)$ cause transitions between $\mathsf{H}^\xi_l$ subspaces with different $l$.

$$\rho'(A_i^\pm) : \mathsf{H}^\xi_l \to \mathsf{H}^\xi_{l\pm 1} : |\eta_K\rangle \mapsto \rho'(A_i^\pm) |\eta_K\rangle$$

This is quite different from the abelian Poincaré case where the generators of the little group only appears in the representations $\sigma'$ on $\mathsf{H}^\sigma$. Here the generators little group appears in both the representations $\sigma'$ on $\mathsf{H}^\sigma$, that we shall discuss shortly, and in these representations $\rho'$ on $\mathsf{H}^\xi$.

Now, these generators may be written in terms of their real generators $\{L_{i,j}, M_{i,j}, Q_i, P_i, I\}$ as given in (3.10) with $Q_i = X_i$ and $P_i = Y_i$. Then with the usual representations for the Heisenberg generators given in (2.6), the matrix elements of $\{L_{i,j}, M_{i,j}\}$ with respect to a coordinate basis are

$$\begin{aligned}
\langle \delta_q | \rho'(Q_i) | \delta_q \rangle &= q^i \\
\langle \delta_q | \rho'(P_i) | \delta_q \rangle &= -i \frac{\partial}{\partial q^i} \\
\langle \delta_q | \rho'(L_{i,j}) | \delta_q \rangle &= -i \frac{1}{s} \left( q^i \frac{\partial}{\partial q^j} - q^j \frac{\partial}{\partial q^i} \right) \\
\langle \delta_q | \rho'(M_{i,j}) | \delta_q \rangle &= \frac{1}{s} \left( q^i q^j - \frac{\partial^2}{\partial q^i \partial q^j} \right)
\end{aligned} \tag{4.5}$$

These expressions are consistent with the representations above noting that

$$\langle \delta_x | \eta_K \rangle = \eta_K(q) = \eta_{k_1,\ldots k_n}(q) = \eta_{k_1}(q^1)\,\eta_{k_2}(q^2)\ldots\eta_{k_n}(q^n)$$

where, as in (2.8), $\eta_{k_a}(x^a)$ are defined in terms of the Hermite functions.

$$\eta_k(s) = \frac{1}{\sqrt{\pi\,2^k\,k!}}\,e^{-\frac{1}{2}s^2}\,H_k(s),\, k\in\mathbb{N},\ s\in\mathbb{R}$$

An equally valid representation is in terms of the momentum basis (2.7)

$$\langle \epsilon_p | \rho'(Q_i) | \epsilon_p \rangle = i\,\frac{\partial}{\partial p^i}$$
$$\langle \epsilon_p | \rho'(P_i) | \epsilon_p \rangle = p^i$$
$$\langle \epsilon_p | \rho'(L_{i,j}) | \epsilon_p \rangle = i\,\frac{1}{s}\left(p^i\,\frac{\partial}{\partial p^j} - p^j\,\frac{\partial}{\partial p^i}\right)$$
$$\langle \epsilon_p | \rho'(M_{i,j}) | \epsilon_p \rangle = \frac{1}{s}\left(p^i\,p^j - \frac{\partial^2}{\partial p^i\,\partial p^j}\right)$$

From this it is clear that the $\rho'(L_{i,j})$ are just the usual differential operator expressions for the $\mathcal{SO}(n)$ generators. These operators do not intrinsically appear in the Poincaré representation theory but do so here do to the nonabelian nature of the Heisenberg normal subgroup.

We now have a basic characterization of the Hilbert spaces of particle states and the representations how the various representations act. We have characterized $\rho'$.

## 4.2 The representation $\sigma'$ of the generators of $\mathcal{U}(n)$.

The final issue is to give the explicit form of the finite dimensional unitary irreducible representations $\sigma'(Z_{i,j})$ of the algebra of $\mathcal{U}(n)$. This problem has been solved both for the representation of the algebra and also for the group. We do not derive the results but simply summarize them here. We will follow the approach of Gel'fand [16,2]. To establish notations and ensure the continuity of this presentation, we summarize a few notational facts.

A convenient basis of $\mathsf{H}^\sigma$ is the Gel'fand basis that is derived from the subgroup chain $\mathcal{U}(n) \supset \mathcal{U}(n-1) \supset \ldots \supset \mathcal{U}(1)$. An element $|\psi\rangle^\sigma$ is indexed by the quantities

$$|\psi\rangle^\sigma = \left| \begin{array}{ccccc} m_{1,n} & m_{2,n} & \ldots & & m_{n,n} \\ & \ldots & & \ldots & \\ & m_{1,2} & & m_{2,2} & \\ & & m_{1,1} & & \end{array} \right\rangle \qquad (4.6)$$

For the group $\mathcal{U}(n)$, the top row $m_{i,n}$ are constant for the representation, and in fact, label the representation uniquely. The remaining $m_{i,j}$, $1 \le i,\ j < n$ label specific states in the representation. It is clear that this repeats itself recursively. For $\mathcal{U}(n-1)$, it is the row $m_{i,n-1}$ that labels the representation and the remaining $m_{i,j}$, $1 \le i,\ j < n-1$ label the states in $\mathcal{U}(n-1)$. This may be repeated recursively until we reach $\mathcal{U}(1)$

Consider $\mathcal{U}(n)$ and $\mathcal{U}(n-1)$. The row $m_{i,n-1}$ that are constants labelling representations of $\mathcal{U}(n-1)$ but are variable state labels in the context of $\mathcal{U}(n)$. Thus multiple representations of $\mathcal{U}(n-1)$ appear in $\mathcal{U}(n)$.

The states are further restricted by the following inequalities

$$m_{i,j} \geq m_{i+1,j}$$
$$m_{i,j} \geq m_{i,j-1} \geq m_{i+1,j}$$

where $i = 1, .j-1$, $j = 1, ..n$. If you put the above inequalities into the triangular form above, a simple pattern emerges.

Now, the requirement of unitarity puts further restrictions on the $m_{i,j}$. In the case of the compact group $\mathcal{U}(n)$ it follows that all the entries are either integral or half integral. It immediately follows that the representations are finite dimensional.

The general form of the unitary operators $\sigma'(Z_{i,j})$ have been determined for the above basis. The first step is to make the above notation more compact. We define

$$|\psi\rangle^\sigma = |M\rangle$$

where $M = [\![m_{i,j}]\!]$ $a = 1, ..b$, $b = 1, n$ is the triangular array given above in (4.6). Furthermore, define $I_{k,l}$ to be a similar triangular array with all entries $m_{i,j} = 0$ except the one entry where $i = k$ and $j = l$, $m_{i,j} = 1$.

The problem is then to determine

$$\sigma'(Z_{i,j})|\psi\rangle^\sigma = \sigma'(Z_{i,j})|M\rangle$$

All of the elements $Z_{a,b}$ may be computed from the elements $Z_{i,i}$ and $Z_{i,i+1}$ using the commutation relations (3.1). Therefore, it is only necessary to determine the representation of these elements. The result is [36]

$$\sigma'(Z_{k,k})|M\rangle = \left(\sum_{i=1}^{k} m_{k,i} - \sum_{i=1}^{k-1} m_{k-1,i}\right)|M\rangle,$$
$$\sigma'(Z_{k,k+1})|M\rangle = \sum_{i=1}^{k} f_{k,i}(M)|M+I_{k,i}\rangle, \qquad (4.7)$$
$$\sigma'(Z_{k+1,k})|M\rangle = \sum_{i=1}^{k} f_{k,i}(M-I_{k,i})|M-I_{k,i}\rangle$$

where

$$f_{k,j}(M) \doteq \frac{\prod_{i=1}^{k+1}\sqrt{m_{k+1,i}-m_{k,j}-i+j}\ \prod_{i=1}^{k-1}\sqrt{m_{k-1,i}-m_{k,j}-i+j-1}}{\prod_{i=1,i\neq j}^{k-1}\sqrt{m_{k,i}-m_{k,j}-i+j-1}\ \sqrt{m_{k,i}-m_{k,j}-i+j}} \qquad (4.8)$$

The matrix elements of the representation of the algebra are then the matrices $\Sigma = [\![\Sigma_{\tilde{M},M}]\!]$ defined by

$$\Sigma_{i,j\ \tilde{M},M} = \langle \tilde{M}| \sigma'(Z_{i,j}) | M \rangle \tag{4.9}$$

Again, $\Sigma$ is a finite dimensional matrix for the compact case $C(n)$.

Note in particular that the eigenvalue of the first order Casimir invariant $\sigma'(D_1) = \sigma'(U)$ defined in (3.6) is

$$\sigma'(U)|M\rangle = \sigma'(\delta^{k,l} Z_{k,l})|M\rangle = d_1 |M\rangle, \tag{4.10}$$

with

$$d_1 = \sum_{i=1}^{n} m_{n,i}$$

The higher order invariants may be similarly computed. $\sigma'(D_2) = \delta^{i,l} \delta^{j,k} \sigma'(Z_{i,j} Z_{k,l})$ is given by

$$\sigma'(D_\alpha)|M\rangle = d_\alpha |M\rangle, \tag{4.11}$$

with

$$d_\alpha = p^\alpha(m_{n,i})$$

where $p^\alpha$ is an $\alpha$ order polynomial in the $m_{n,i}$, $i = 1, ..n$. In particular, for $\mathcal{U}(2)$,

$$d_2 = m_{2,1} + m_{2,1}^2 - m_{2,2} + m_{2,2}^2$$

and for $\mathcal{U}(3)$,

$$d_2 = m_{3,1}^2 + 2 m_{3,1} + m_{3,2}^2 - 2 m_{3,3} + m_{3,3}^2$$
$$d_3 = m_{3,1}^3 + 4 m_{3,1}^2 + 4 m_{3,1} + m_{3,2}^3 + m_{3,2}^2 -$$
$$2 m_{3,2} m_{3,2} m_{3,3} - m_{3,1} m_{3,2} - m_{3,1} m_{3,3} + -2 m_{3,3} - 2 m_{3,3}^2 + m_{3,3}^3$$

Higher order unitary group Casimir invariants $d_\alpha$, $\alpha = 3, 4..n$ may similarly be computed. This completely specifies the representations albeit in a somewhat computational opaque form. Calculations beyond $n = 1$ or 2 rapidly require the symbolic computational capabilities of *Mathematica*.

Note in particular that if the representation of $\mathcal{U}(n) = \mathcal{U}(1) \otimes \mathcal{SU}(n)$ is such that the representation of $\mathcal{SU}(n)$ is trivial with $d_2 = 0, d_3 = 0, ...d_n = 0$, then the matrix elements are simply

$$\Sigma_{i,j\ \tilde{M},M} = d_1 \delta_{i,j} \delta_{\tilde{M},M} \tag{4.12}$$

### 4.3 Casimir operator field equations

We now know explicitly the general form of the representations $\varrho'$ of the algebra of $C(n)$ with respect to both a co-ordinate and coherent basis of $\{Z_{i,j}, A_i^{\pm}, I\}$. We can use this immediately to compute the representations $\varrho'(W_{i,j})$ of the generators $W_{i,j}$ used to define the Casimir invariants.

$$\varrho'(I)|\psi\rangle = c_1|\psi\rangle$$
$$\varrho'(W_{i,j})|\psi\rangle = \varrho'(A_i^+ A_j^-)|\psi\rangle - \varrho'(Z_{i,j})c_1|\psi\rangle$$

Now, using (3.14), this becomes

$$\varrho'(W_{i,j})|\psi\rangle = |\psi\rangle^\sigma \otimes \rho'(A_i^+ A_j^- - I Z_{i,j})|\psi\rangle^\xi - c_1 \sigma'(Z_{i,j})|\psi\rangle^\sigma \otimes |\psi\rangle^\xi$$

The matrix elements of the representations of $\sigma'(Z_{i,j})$ are just the finite dimensional matrices $\Sigma_{i,j\, \tilde{M},M} = \langle \tilde{M}|\sigma'(Z_{i,j})|M\rangle$ with $\{|M\rangle\}$ a basis of $\mathcal{H}^\sigma \simeq \mathfrak{V}^N$, $N = \dim \sigma$.

4.3.1 Representation of the second order Casimir invariant operator

The matrix elements of $\rho'_l(W_{i,j})$ in a co-ordinate basis $|\delta_q\rangle$ of $\mathcal{H}^\xi$ follow directly from (4.5)

$$\langle \delta_q|\rho'(W_{i,j})|\delta_q\rangle = \rho'(Z_{i,j}) = (1 + \tfrac{c_1}{s})\xi'(A_i^+)\xi'(A_j^-) =$$
$$\tfrac{1}{2}(1 + \tfrac{c_1}{s})\langle \delta_q|\xi'((Q_i + i P_i))\xi'((Q_j - i P_i))|\delta_q\rangle$$
$$= a\langle \delta_q|\xi'(Q_i)\xi'(Q_j) + \xi'(P_i)\xi'(P_j)|\delta_q\rangle$$
$$= a\left(q^i q^j - \tfrac{\partial^2}{\partial q^i \partial q^j}\right)$$

where $a = \tfrac{1}{2}(1 + \tfrac{c_1}{s})$. Combining this with the finite dimensional matrix elements of (4.9), we have the full representation

$$\langle \delta_q, \tilde{M}|\varrho'(W_{i,j})|\delta_q, M\rangle = \left(a\left(q^i q^j - \tfrac{\partial^2}{\partial q^i \partial q^j}\right)\delta_{\tilde{M},M} - c_1 \Sigma_{i,j\, \tilde{M},M}\right)$$

In particular, using (4.10)

$$\langle \delta_q, \tilde{M}|\varrho'(W_{i,i})|\delta_q, M\rangle = \left(a\left(q^2 - \tfrac{\partial^2}{\partial q^2}\right) - c_1 m\right)\delta_{\tilde{M},M} \qquad (4.13)$$

with $q^2 = \delta_{i,j} q^i q^j$. We now can compute the Casimir field equations in the position co-ordinate basis. The co-ordinate representation of the field equations is

$$\langle \delta_q, M|\varrho'(C_\alpha)|\psi\rangle = c_\alpha \langle \delta_q, M|\psi\rangle$$

That is

$$\langle \delta_q, M|\varrho'(C_\alpha)|\delta_q, \tilde{M}\rangle \psi_{\tilde{M}}(q) = c_\alpha \psi_M(q)$$

The first equation $\alpha = 1$ for which $C_1 = I$ is simply

$$\langle \delta_q, \mathrm{M} | \varrho'(I) | \delta_q, \tilde{\mathrm{M}} \rangle \psi_{\tilde{\mathrm{M}}}(q) = c_1 \psi_{\mathrm{M}}(q)$$

For the second Casimir field equation $\alpha = 2$, note $C_2 = W = W_{i,i}$ and so immediately from (4.13) that

$$\left( q^2 - \frac{\partial^2}{\partial q^2} - \frac{c_2 + c_1 d_1}{a} \right) \psi_{\mathrm{M}}(q) = 0$$

This of course is the familiar Harmonic oscillator equation with solution

$$\psi_{l,\mathrm{M}}(q) = \eta_l(q) = \eta_{k_1}(q^1) \eta_{k_2}(q^2) \ldots \eta_{k_n}(q^n) \quad , l = \sum_{i=1}^{n} k_i$$

where the functions $\eta_k(s)$ are defined in (2.8) as

$$\eta_k(s) = \frac{1}{\sqrt{\pi} \, 2^k \, k!} \, e^{-\frac{1}{2} s^2} H_k(s), \, k \in \mathbb{Z}^+, \, s \in \mathbb{R}$$

provided that $2l + n = \frac{c_2 + c_1 d_1}{a}$ with $l$ is a non negative integer and $c_1, c_2$ are independent of $l \in \mathbb{Z}^+$ and $d_1 \in \mathbb{Z}$ as it is constant for the unitary irreducible representation. Note $a$ is a constant defining the projective representation. This is satisfied if

$$c_1 d_1 = 2 a l$$
$$c_2 = n a, \quad a = \frac{1}{2} \left( 1 + \frac{c_1}{s} \right)$$

Solving these gives

$$a = \frac{s}{2(s-1)}, \quad c_1 = 2a, \quad c_2 = na, \quad d_1 = l \tag{4.14}$$

As an aside, it is clear that this is only valid for a projective representation with $s \neq 1$, and is not valide for an ordinary representation with $s = 1$ This leads to the key result that the boundary conditions pick out a specific projective representation constant and second Casimir invariant is defined by the projective constant and the dimensionality of the space.

Note that the $l$ label the $\mathsf{H}_l^\xi$ subspaces of $\mathsf{H}^\xi$ of the $\rho$ representation discussed above and $d_1$ is the first Casimir invariant of $\mathcal{U}(n) = \mathcal{U}(1) \otimes \mathcal{SU}(n)$ in the $\sigma$ representations on $\mathsf{H}^\sigma$. The identification $l = d_1$ requires that in the full representation $\varrho$, only these representations appear in $\varrho = \sigma \otimes \rho$ acting on $\mathsf{H}^\sigma \otimes \mathsf{H}_l^\xi$. This is where the *internal* $\sigma$ and the *external* position degrees of freedom couple.

It follows from this that any representation of $\mathcal{SU}(n)$ in the above decomposition $\mathcal{U}(n)$ will satisfy the equation as there are not constraints on the higher order Casimir invariants. In particular the *scalar* case where the representation of $\mathcal{SU}(n)$ is trivial is a solution. This is the usual scalar harmonic oscillator and parallels the Klein-Gordon equation with zero spin, or trivial forth order Casimir invariant. Of course it is necessary to establish that this *scalar* case, in fact, satisfies the higher order Casimir invariant equations as was also the case in the Poincaré theory.

We could have done the calculation instead with respect to a basis $|\epsilon_p\rangle$ in which momentum is diagonal yielding

$$\left(p^2 - \frac{\partial^2}{\partial p^2} - (2l+n)\right)\tilde{\psi}_l(p) = 0$$

The solution is again $\tilde{\psi}_l(p) = \eta_l(p)$. Defining $\mu^2 = p^2 > 0$, this may be composed also in polar coordinates as

$$\left(\nabla_\theta^2 - j(j+n-1)\right)\psi(\theta_1, ... \theta_{n-1}) = 0, l \in \mathbb{Z} \geq 0,$$

$$\left(-\frac{\partial^2}{\partial \mu^2} - \frac{n}{\mu}\frac{\partial}{\partial \mu} + \frac{j(j+n-1)}{\mu^2} + \mu^2 - (2l+n)\right)\psi(\mu) = 0$$

$\nabla_\theta^2$ is the Laplacian on $\mathbb{S}^{n-1}$. This is the precursor for the mass calculation in the following section. This has the solution in terms of spherical functions $Y^j_{m_1..m_{n-1}}$ and Legendre polynomials $L^j_k$ [7].

In the physical interpretation as the time independent harmonic oscillator, the total energy is an eigenvalue with values $(2l+n)$ and the angular momentum has eigenvalues $j(j+n-1)$. However, the precursor to mass in this Euclidean theory, $\mu$ is not diagonal. Rather the Euclidean mass is statistically determined by the probability density $\psi^\dagger(\mu)\mu\psi(\mu)$ with $\psi(\mu)$ appropriately normalized. We will return to this in the $C(1, n)$ theory.

4.3.2 Representation of the fourth order Casimir invariant operator

The third equation $\alpha = 4$ for which $C_4 = W^2 = W_{i,j}W_{j,i}$ is

$$\left(a\left(q^i q^j - \frac{\partial^2}{\partial q^i \partial q^j}\right)\delta_{\tilde{M},M} - c_1 \Sigma_{i,j\,\tilde{M},M}\right)$$
$$\left(a\left(q^j q^i - \frac{\partial^2}{\partial q^j \partial q^i}\right)\delta_{\tilde{M},M} - c_1 \Sigma_{j,i\,\tilde{M},M}\right)\psi_{\tilde{M}}(q) = c_4 \psi_M(q)$$

This process may be repeated until the full set $\alpha = 1, 2, 4, 6, ... 2n$ is obtained for the $n + 1$ independent Casimir invariants of $C(n)$. Clearly these equations rapidly become rather complex.

To gain further insight, note that with straightforward Lie algebraic rearrangement, $C_4$ may be cast in the form (3.7). The corresponding field equation that is the eigenvalue equation of the representation $\varrho'(C_4)$ for this form of $C_4$ is

$$\Sigma_{i,j\,M,\tilde{M}}\left(q^i - \frac{\partial}{\partial q^i}\right)\left(q^j + \frac{\partial}{\partial q^j}\right)\psi_{\tilde{M}}(q) =$$
$$\sum_{\alpha=1}^{2}\sum_{\kappa=0}^{2-\alpha+1} f^2_{\kappa,\alpha}(n, c_1, c_2, c_4) d_\alpha{}^\kappa \psi_M(q) \quad (4.15)$$

Using (3.8), as $c_1$ and $c_2$ have been determined above in (4.14), the nonzero $f^2_{\kappa,\alpha}(n, c_1, c_2, c_4)$ are

$$\begin{aligned} f^2_{0,1}(n, C_1, C_2) &= n - n^2(\tfrac{1}{s^2} - \tfrac{1}{4}) - \tfrac{1}{2} C_4 \\ f^2_{1,1}(n, C_1, C_2) &= (-2 + \tfrac{n}{s} - \tfrac{3n}{2}) \\ f^2_{2,1}(n, C_1, C_2) &= 2(1 + \tfrac{1}{n}) \\ f^2_{2,2}(n, C_1, C_2) &= \tfrac{1}{2} \end{aligned} \qquad (4.16)$$

Note that these constants depend only on the dimension of the group and the Casimir invariants of the representation. Note that for this invariant, the internal $\mathcal{U}(n))$ spin degrees of freedom couple. This can be repeated for higher order equations. In each case, (at least up to $C_\beta$ with $\beta = 8$) it reduces the eigenvalue to a second order partial differential equation coupling with powers the $\Sigma_{i,j\,\mathrm{M},\tilde{\mathrm{M}}}$ and a polynomial in the Casimir invariants of $\mathcal{U}(n)$, $d_\alpha$, $\alpha = 1, ..\beta$ with coefficients depending on the Casimir invariants of the quaplectic group $c_\alpha$, $\alpha = 1, ..\beta$, the projective constant $s$ and the dimension $n$ of the theory.

Thus, it is clear that even though $C(n)$ is a dynamical group on the phase space, it still yields partial differential field equations which are analogous to those that arise from the Euclidian $\mathcal{E}(n)$ and Poincaré groups $\mathcal{E}(1, n)$. The mathematical formalism is precisely the same, just a different dynamical group.

### 4.3.3 Coherent basis

The equations can be cast into a remarkably simpler form by considering a coherent basis $|\eta_K, \mathrm{M}\rangle$ instead of a position coordinate basis.

The matrix elements are the finite dimensional matrices in terms of the coherent basis

$$\langle \tilde{\mathrm{M}}, \eta_{\tilde{K}} | \varrho'(W_{i,j}) | \eta_K, \mathrm{M} \rangle = a\, \Xi_{i,j,\tilde{K},K}\, \delta_{\tilde{\mathrm{M}},\mathrm{M}} - c_1\, \delta_{\tilde{K},K}\, \Sigma_{i,j\,\tilde{\mathrm{M}},\mathrm{M}}$$

We have noted that the Hilbert space $\mathsf{H}^\xi$ may be decomposed into a countably infinite ladder of finite dimensional vector spaces $\mathfrak{V}^l$, $l = 1, 2. ....$. The action of the generators $\rho'(Z_{i,j})$ maps these spaces into themselves. However, the action of $\rho'(A^\pm_i)$ causes transitions $l \pm 1$. in general the $\Xi_{i,j,\tilde{K},K}$ are countably infinite dimensional matrix.

Note that the product $\rho'(A^+_i)\,\rho'(A^-_i)$ leaves these spaces invariant and so

$$\rho'_l(W_{i,j}) : \mathsf{H}^\xi_l \to \mathsf{H}^\xi_l \text{ where } \mathsf{H}^\xi_l \simeq \mathfrak{V}^l$$

where $\rho'_l$ is $\rho'_l$ restricted to $\mathfrak{V}^l$. As the Casimir invariants are determined entirely by the $W_{i,j}$, this is a very remarkable property as it decouples the field equations from an infinite dimensional equation to a ladder of finite dimensional equations. That is

$$\varrho'_l(W_{i,j}): \mathfrak{V}^N \otimes \mathfrak{V}^l \to \mathfrak{V}^N \otimes \mathfrak{V}^l \quad \text{where } \mathsf{H}^\xi_l \simeq \mathfrak{V}^l \text{ and } \mathsf{H}^\sigma \simeq \mathfrak{V}^N, \, l = 1, 2, .. \infty$$

Thus the representations $\Xi^l_{i,j,\tilde{K},K}$ restricted to $\mathsf{H}^\xi_l$ are just finite dimensional matrices. Thus

$$\langle \tilde{M}, \eta_{\tilde{K}} | \varrho'_l(W_{i,j}) | \eta_K, M \rangle = a \, \Xi^l_{i,j,\tilde{K},K} \, \delta_{\tilde{M},M} - c_1 \, \delta_{\tilde{K},K} \, \Sigma_{i,j \, \tilde{M},M}$$

are finite dimensional for each $l$. The Casimir invariants are independent of $l$ and so any non trivial $l$ may be chosen for the calculation, say $l = 2$. The eigenvalue equations may be constructed a powers of these finite matrices and the eigenvalues and eigenvectors solved.

# 5: The field equations of the group $C(1, n)$

The general case of $C(1, n)$ follows the $C(n)$ discussion with the key difference that the little group $\mathcal{U}(1, n)$ is noncompact and therefore the representation $\sigma$ are countably infinite rather than finite dimensional representations, $\mathsf{H}^\sigma \simeq \mathfrak{V}^\infty$.

The steps are the same. First determine the unitary irreducible representations $\rho'$ of the generators $\{I, A_a^\pm, Z_{a,b}\}$ with $a, b = 0, 1, \ldots n$ of $C(1, n)$ on $\mathsf{H}^\xi \simeq \mathcal{L}^2(\mathbb{R}^{n+1}, \mathbb{R})$. Next, the standard countably infinite dimensional representations $\sigma'$ of $\mathcal{U}(1, n)$ on the vector space $\mathsf{H}^\sigma \simeq \mathfrak{V}^\infty$ are summarized. These representations can be decomposed $\mathcal{U}(1, n) \supset \mathcal{U}(n) \supset \mathcal{U}(n-1) \supset \ldots \supset \mathcal{U}(1)$ and so the representations of $\mathcal{U}(1, n)$ may be viewed as a countably infinite sum or ladder of finite dimensional representations of $\mathcal{U}(n)$ [30,36]. Finally, these are combined to give the full representation $\varrho'$ of the generators $\{I, A_a^\pm, Z_{a,b}\}$ of $C(1, n)$ on $\mathsf{H}^\varrho \simeq \mathfrak{V}^\infty \otimes \mathcal{L}^2(\mathbb{R}^{n+1}, \mathbb{R})$. These may be used to directly compute the representations $\varrho'$ of the generators $W_{a,b} \doteq A_a^+ A_b^- - I Z_{a,b}$. These generators may be used to construct the Casimir invariants $C_1 = I$, $C_2 = W$, $C_4 = W^2, \ldots C_{2(N_c-1)} = W^{N_c-1}$ using (3.4). The field equations are then the $N_c = n + 2$ equations

$$\varrho'(I) |\psi\rangle = c_1 |\psi\rangle$$
$$\varrho'(W^\alpha) |\psi\rangle = c_\alpha |\psi\rangle \quad \alpha = 1, 2, 4, \ldots 2(n+1)$$

with $|\psi\rangle \in \mathsf{H}^\varrho$.

## 5.1 The representation $\rho'$ of the generators of $C(1, n)$.

Following the $C(n)$ analysis, the representations $\rho'$ of the algebra of the full group $C(1, n)$ act on the Hilbert space $\mathsf{H}^\xi \simeq \mathcal{L}^2(\mathbb{R}^{n+1}, \mathbb{R})$ of the representations $\xi$ of the normal subgroup $\mathcal{H}(1, n)$. The representations $\rho'$ are the extension to the full set of generators $\{Z_{a,b}, A_i^\pm, I\}$, $a, b, \ldots = 0, 1, \ldots n$ of $C(1, n)$ given by

$$\rho'(I) = \xi'(I)$$
$$\rho'(A_a^\pm) = \xi'(A_a^\pm)$$
$$\rho'(Z_{a,b}) = \tfrac{-1}{s} \xi'(A_a^+) \xi'(A_b^-)$$

where $s$ is the projective constant. A basis for the Hilbert space $\mathsf{H}^\xi$ is given in (2.8) as $|\eta_K\rangle = |\eta_{k_0, k_1, \ldots k_n}\rangle$ with $K = (k_0, k_1, \ldots k_n)$.

Defining $I_a$ now to be an $n + 1$ tuple of 0's with a 1 at the $a$ th position, $I_a = (0, 0, , , 1, \ldots 0)$, this may be written more compactly as

$$\xi'(A_0^-)|\eta_K\rangle = \sqrt{k_0+1}\,|\eta_{K+I_0}\rangle$$
$$\xi'(A_0^+)|\eta_K\rangle = \sqrt{k_0}\,|\eta_{K-I_0}\rangle$$

$$\xi'(A_i^+)|\eta_K\rangle = \sqrt{k_i+1}\,|\eta_{K+I_i}\rangle$$
$$\xi'(A_i^-)|\eta_K\rangle = \sqrt{k_i}\,|\eta_{K-I_i}\rangle$$

where as always $i, j = 1, 2, ..n$. Note that $\xi'(A_0^\pm)$ *acts like* $\xi'(A_i^\mp)$. It follows directly that

$$\rho'(Z_{i,0})|\eta_K\rangle = \tfrac{1}{s}\sqrt{(k_0+1)(k_i+1)}\,|\eta_{K+I_i+I_0}\rangle$$
$$\rho'(Z_{0,i})|\eta_K\rangle = \tfrac{1}{s}\sqrt{k_i k_0}\,|\eta_{K-I_i-I_0}\rangle$$
$$\rho'(Z_{j,i})|\eta_K\rangle = \tfrac{1}{s}\sqrt{(k_j+1)k_i}\,|\eta_{K-I_i+I_j}\rangle, i \neq j$$
$$\rho'(Z_{a,a})|\eta_K\rangle = \tfrac{1}{s}k_a|\eta_K\rangle$$

It may be verified that the representation satisfies the commutation relations (3.1) for the algebra of $C(1, n)$ as Hermitian operators on the Hilbert space $\mathsf{H}^\xi$ and so are, indeed, representations of the algebra. Again, these are degenerative representations with a single independent Casimir invariant.

The little group may be factored as $\mathcal{U}(1, n) = \mathcal{U}(1) \otimes \mathcal{SU}(1, n)$. Note that the generator $U$ of the algebra of $\mathcal{U}(1)$ is defined by $U = \eta^{a,b} Z_{a,b}$ and it follows immediately that the representation is

$$\rho'(Y)|\eta_K\rangle = (-k_0 + k_1 + k_2 + ... k_n)|\eta_K\rangle = k|\eta_K\rangle \qquad (5.1)$$

with $k = -k_0 + \sum_{i=1}^n k_i$ where again $k_a \in \mathbb{Z} \geq 0$. The negative appearing as the coefficient of $k_0$ means that there are now an infinite number of combinations of the $k_a$ that satisfy this constraint. Define, as before, $l = \sum_{i=1}^n k_i$, and the $l$ label the $\mathcal{U}(n)$ invariant subspaces $\mathsf{H}_l^\xi \simeq \mathfrak{V}^l$. Then for each $l$, there is a $k_0$ satisfying this equation $k = l - k_0$. The subspaces $\mathcal{U}(1, n)$ invariant subspaces labelled by $k$ are countably infinite dimensional

Thus, the $\mathsf{H}_l^\xi$ are now countably infinite dimensional $\mathsf{H}_l^\xi \simeq \mathfrak{V}^\infty$. These, in turn may be decomposed into the finite dimensional $\mathcal{U}(n)$ invariant subspaces, which again may appear with multiplicity

$$\mathsf{H}^\xi = \oplus_{k=1}^\infty \mathsf{H}_k^\xi = \oplus_{k=1}^\infty \oplus_{l=k}^\infty \mathfrak{V}^l$$

The subspaces $\mathsf{H}_l^\xi$ are invariant under the generators of $\mathcal{U}(1, n)$

$$\rho'(Z_{a,b}): \mathsf{H}_l^\xi \to \mathsf{H}_l^\xi : |\eta_K\rangle \mapsto \rho'(Z_{a,b})|\eta_K\rangle$$

The representations of the generators $\rho'(A_a^\pm)$ of $\mathcal{H}(1, n)$ cause transitions between $\mathsf{H}_k^\xi$ subspaces with different $k$.

$$\rho'(A_i^\pm): \mathsf{H}_k^\xi \to \mathsf{H}_{k\pm 1}^\xi \ , \ \rho'(A_0^\pm): \mathsf{H}_k^\xi \to \mathsf{H}_{k\mp 1}^\xi$$

Now, these generators may be written in terms of their real generators $\{J_{i,j}, M_{i,j}, N_i, K_i, R, Q_i, P_i, T, E, I\}$ as given in (3.13). These could be written in the four notation (3.10) but this enables us to only consider the equations involving $\{N_i, K_i, R, T, E\}$ in addition to the $\{J_{i,j}, M_{i,j}, Q_i, P_i, I\}$ considered in the previous section.

The real representations of the algebra of $C(1, n)$ are the representations of the generators of $C(n)$ given in (4.5) plus the representations of the time and energy generators $\{T, E\}$ and the velocity and momentum boosts $\{K_i, N_i\}$, which, with respect to position time coordinates, are

$$\langle \delta_t | \rho'(T) | \delta_t \rangle = t$$
$$\langle \delta_t | \rho'(E) | \delta_t \rangle = -i \frac{\partial}{\partial t}$$
$$\langle \delta_t \delta_q | \rho'(K_i) | \delta_t \delta_q \rangle = i \left( t \frac{\partial}{\partial q^j} + q^j \frac{\partial}{\partial t} \right)$$
$$\langle \delta_t \delta_q | \rho'(N_i) | \delta_t \delta_q \rangle = t\, q^i - \frac{\partial}{\partial q^j} \frac{\partial}{\partial t}$$

These expressions are consistent with the representations above noting that

$$\langle \delta_{t,q} | \eta_K \rangle = \eta_K(t, q) = \eta_{k_0, k_1, \ldots k_n}(q) = \eta_{k_0}(t)\, \eta_{k_1}(q^1)\, \eta_{k_2}(q^2) \ldots \eta_{k_n}(q^n)$$

where, as in (2.8), $\eta_{k_a}(x^a)$ are defined in terms of the Hermite functions (2.8). Note that the exponential appearing in the expression for $\eta_K(t, q)$ is $e^{-\frac{1}{2}(t^2+q^2)}$ and not $e^{-\frac{1}{2}(-t^2+q^2)}$ as one might expect at first glance.

An equally valid representation is in terms of the time momentum basis

$$\langle \delta_t \epsilon_p | \rho'(K_i) | \delta_t \epsilon_p \rangle = \left( t\, p^j - \frac{\partial}{\partial p^j} \frac{\partial}{\partial t} \right)$$
$$\langle \delta_t \epsilon_p | \rho'(N_i) | \delta_t \epsilon_p \rangle = -i \left( t \frac{\partial}{\partial p^j} + p^j \frac{\partial}{\partial t} \right)$$

There are similar representations with respect to the energy momentum basis $|\epsilon_e \epsilon_p\rangle$ and the energy position basis $|\epsilon_e \delta_q\rangle$.

We now have a basic characterization of the Hilbert spaces of particle states and the representations how the various representations act. We have characterized $\rho'$.

## 5.2 The representation $\sigma'$ of the generators of $\mathcal{U}(1, n)$.

The next task is to give the explicit form of the representations $\sigma'(Z_{a,b})$ of the algebra of $\mathcal{U}(1, n)$. This problem has been solved both for the representation of the algebra and also for the group in [30,36]. This is briefly summarized in the notation of this paper as follows.

arXiv:math-ph/0404077 Canonically Relativistic Quantum MechanicsA convenient basis of $\mathsf{H}^\sigma$ is the Gel'fand basis that is derived from the subgroup chain $\mathcal{U}(1, n) \supset \mathcal{U}(n) \supset \mathcal{U}(n-1) \supset ... \supset \mathcal{U}(1)$. An element $|\psi\rangle^\sigma$ is indexed by the quantities

$$|\psi\rangle^\sigma = \left| \begin{array}{cccccc} m_{1,n+1} & m_{2,n+1} & & ... & m_{n,n+1} & m_{n+1,n+1} \\ & m_{1,n} & m_{2,n} & ... & m_{n,n} & \\ & & ... & ... & & \\ & & m_{1,2} & m_{2,2} & & \\ & & & m_{1,1} & & \end{array} \right\rangle$$

This is the same labelling of states that was given for $\mathcal{U}(n)$ with a new row added for the $n+1$ degree of freedom. This is where we need to make a notational note. We consistently use $i, j = 1, ..n$ to label the *compact* $n$ degrees of freedom. When adding the noncompact degree of freedom, we generally use indices $a, b = 0, 1, ..n$. labelling the $n+1$ degree of freedom with a $0$. In this section, in order for the required inequalities to take their simplest form, we will label this extra degree of freedom with $a, b = 1, 2, ...n+1$ instead of $0$. The clear equivalence may be used to map between the $a, b = 0, 1..n$ notation of the remainder of the paper.

For the group $\mathcal{U}(1, n)$, the top row $m_{i,n+1}$ are constant for the representation, and in fact, label the representation uniquely. The remaining $m_{i,j}$, $1 \leq i, j \leq n$ label specific states in the representation. It is clear that this repeats itself recursively. For $\mathcal{U}(n)$, it is the row $m_{i,n}$ that labels the representation and the remaining $m_{i,j}$, $1 \leq i, j \leq n-1$ label the states in $\mathcal{U}(n)$.

Consider $\mathcal{U}(n)$ and $\mathcal{U}(n-1)$. The row $m_{a,n-1}$ that are constants labelling representations of $\mathcal{U}(n-1)$ but are variable state labels in the context of $\mathcal{U}(n)$. Thus multiple representations of $\mathcal{U}(n-1)$ appear in $\mathcal{U}(n)$.

The states are further restricted by the following inequalities

$$m_{a,b} \geq m_{a+1,b}$$
$$m_{a,b} \geq m_{a,b-1} \geq m_{a+1,b}$$

where $a = 1, .b-1$, $b = 1, ..n+1$. If you put the above inequalities into the triangular form above, a simple pattern emerges.

Now, the requirement of unitarity puts further restrictions on the $m_{a,b}$. In the case of the compact group $\mathcal{U}(n)$ it follows that all the entries are either integral or half integral. It immediately follows that the representations are finite dimensional.

The unitarity conditions for the non compact case $\mathcal{U}(1, n)$ is significantly more complex as described in [36]. Suffice it to say that in certain cases, the entries $m_{a,n+1}$ are not necessarily bounded for the noncompact case and so the representations become countably infinite dimensional. These, considerations, of course, determine the exact nature of the Hilbert space $\mathsf{H}^\sigma$.

Consider $\mathcal{U}(n)$ and $\mathcal{U}(n-1)$. The row $m_{i,n-1}$ that are constants labelling representations of $\mathcal{U}(n-1)$ but are variable state labels in the context of $\mathcal{U}(n)$. Thus multiple representations of $\mathcal{U}(n-1)$ appear in $\mathcal{U}(n)$. Now for $\mathcal{U}(1,n)$ and $\mathcal{U}(n)$ the same is true, except, as noted above, in certain representations the $m_{i,j}$ in general and the $m_{i,n}$ in particular may be unbounded and hence a infinite set of representations of $\mathcal{U}(n)$ may appear in representations of $\mathcal{U}(1,n)$.

The general form of the unitary operators $\sigma'(Z_{a,b})$ have been determined for the above basis. Again, define

$$|\psi\rangle^\sigma = |\mathrm{M}\rangle$$

where $\mathrm{M} = [\![m_{a,b}]\!]$ $a = 1,..b$, $b = 1, n+1$ is the triangular array given above. Furthermore, define $\mathrm{I}(c,d)$ to be a similar triangular array with all entries $m_{a,b} = 0$ except the one entry where $a = c$ and $b = d$, $m_{a,b} = 1$.

The problem, as in the compact case, is then to determine the matrix elements.

$$\sigma'(Z_{a,b})|\psi\rangle^\sigma = \sigma'(Z_{a,b})|\mathrm{M}\rangle$$

Again, only the $Z_{a,a}$ and $Z_{a,a\pm 1}$ are required as the commutation relations for the quaplectic group can be used to compute the remainder. The result restricted to the $C(n)$ subgroup, $Z_{i,i}$ and $Z_{i,i\pm 1}$ is given in (4.7). The additional generators are

$$\sigma'(Z_{0,0})|\mathrm{M}\rangle = -\left(\sum_{i=1}^{n+1} m_{k,i} - \sum_{i=1}^{n} m_{k-1,i}\right)|\mathrm{M}\rangle,$$

$$\sigma'(Z_{0,1})|\mathrm{M}\rangle = i\sum_{i=1}^{n+1} f_{k,i}(\mathrm{M})|\mathrm{M} + \mathrm{I}_{k,i}\rangle,$$

$$\sigma'(Z_{1,0})|\mathrm{M}\rangle = i\sum_{i=1}^{n+1} f_{k,i}(\mathrm{M} - \mathrm{I}_{k,i})|\mathrm{M} - \mathrm{I}_{k,i}\rangle$$

The matrix elements of the representation of the algebra are then the countably infinite dimensional matrices $\Sigma = [\![\Sigma_{\tilde{\mathrm{M}},\mathrm{M}}]\!]$ defined by

$$\Sigma_{a,b\ \tilde{\mathrm{M}},\mathrm{M}} = \langle\tilde{\mathrm{M}}|\sigma'(Z_{a,b})|\mathrm{M}\rangle$$

Again, $\mathcal{U}(1,n)$ may be decomposed into the direct product $\mathcal{U}(1,n) = \mathcal{U}(1) \otimes \mathcal{SU}(1,n)$. The Casimir invariant $d_1$ is the invariant associated with $\mathcal{U}(1)$

$$d_1 = \sum_{a=0}^{n} \eta^{a,a} m_{a,n+1} = m_{1,n+1} + m_{2,n+1} ..... m_{n,n+1} - m_{n+1,n+1}$$

with the identification $m_{0,n+1} \simeq m_{n+1,n+1}$ and so if the representation of $\mathcal{SU}(1,n)$ is trivial

$$\Sigma_{a,b\ \tilde{M},M} = d_1\ \eta_{a,b}\ \delta_{\tilde{M},M}$$

This completes the specification the representations

## 5.3  Casimir operator field equations

These results may be used immediately to compute the representations $\varrho'(W_{a,b})$ of the generators $W_{a,b}$ used to define the Casimir invariants. Following the same arguments as the compact case, this is

$$\varrho'(W_{a,b})|\psi\rangle = |\psi\rangle^\sigma \otimes \rho'(A_a^+ A_b^- - I\, Z_{a,b})|\psi\rangle^\xi - c_1\, \sigma'(Z_{a,b})|\psi\rangle^\sigma \otimes |\psi\rangle^\xi$$

There are 4 natural *coordinate* representatives depending on whether we diagonalize position-time, momentum-time, momentum-energy or position-energy. The position-time coordinates are, of course, most familiar and so we begin with this form. It is important to not that there is nothing intrinsic in the formalism that biases us to this diagonalization vs the other three.

### 5.3.1  Representation of the second order Casimir invariant

For this section, we identify $\{X_a\} \simeq \{T, Q_i\}$ and $\{x^a\} \simeq \{t, q^i\}$

The matrix elements of $\rho'_l(W_{a,b})$ in a co-ordinate basis $|\delta_x\rangle$ of $\mathsf{H}^\xi$ are

$$\langle \delta_x|\, \rho'(W_{a,b})\, |\delta_x\rangle = a\!\left(x^a x^b - \frac{\partial^2}{\partial x^a\, \partial x^b}\right)$$

Combining this with the matrix elements of the $\sigma'$ representation, we have the full representation

$$\langle \delta_x, \tilde{M}|\, \varrho'(W_{a,b})\, |\delta_x, M\rangle = \left(a\!\left(x^a x^b - \frac{\partial^2}{\partial x^a\, \partial x^b}\right) \delta_{\tilde{M},M} - c_1\, \Sigma_{i,j\ \tilde{M},M}\right)$$

The matrices $\Sigma_{i,j}$ are now countably infinite dimensional.

We now can compute the Casimir field equations in the position-time coordinate basis. The co-ordinate representation of the field equations is

$$\langle \delta_x, M|\, \varrho'(C_\alpha)\, |\delta_x, \tilde{M}\rangle\, \psi_{\tilde{M}}(q) = c_\alpha\, \psi_M(x)$$

The first equation $\alpha = 1$ for which $C_1 = I$ is simply

$$\langle \delta_x, M|\, \varrho'(I)\, |\delta_x, \tilde{M}\rangle\, \psi_{\tilde{M}}(q) = c_1\, \psi_M(x)$$

For the second equation $\alpha = 2$. Note $C_2 = Y = \eta^{a,b} W_{a,b}$. It follows that $\langle M | \sigma'(U) | \tilde{M} \rangle = d_1 \delta_{M,\tilde{M}}$ is the first Casimir invariant of $\mathcal{U}(n)$ and therefore this equation is

$$\left(x^2 - \frac{\partial^2}{\partial x^2} - \frac{c_2 + c_1 d_1}{a}\right)\psi_M(x) = 0$$

with $x^2 = \eta_{a,b}\, x^a x^b$. Setting $\lambda = \frac{c_2 + c_1 d_1}{a}$, this may be separated in Cartesian co-ordinates as

$$\left((x^i)^2 - \frac{\partial^2}{\partial (x^i)^2} - \lambda_i\right)\psi(x) = 0, \quad \left((x^0)^2 - \frac{\partial^2}{\partial (x^0)^2} + \lambda_0\right)\psi(x) = 0$$

and so $\lambda_i = 2 k_i + 1$ with $k_i \in \mathbb{Z}^+$ and $\lambda_0 = -(2 k_0 - 1)$ with $k_0 \in \mathbb{Z}^+$ where we have inserted the negative to keep $k_0$ a positive integer. Then, as $\lambda = \lambda_0 + \lambda_1 + .. \lambda_n$ we have

$$\lambda = 2(-k_0 + k_1 + ... k_n) + n - 1, \, k_a \in \mathbb{Z}^+$$

This the relativistic Harmonic oscillator equation with $2k + n - 1 = -\frac{c_2 + c_1 d_1}{a}$ with solutions

$$\psi_M(x) = \eta_k(x) = \eta_{k_0}(x^0)\, \eta_{k_1}(x^1)\, \eta_{k_2}(x^2) ... \eta_{k_n}(x^n)$$

Then, as in the compact case, noting $\lambda = 2k + n - 1 = -\frac{c_2 + c_1 d_1}{a}$ this is satisfied if $k = d_1$ and

$$c_1 = 2a$$
$$c_2 = (n-1)a$$

where, again,

$$a = \frac{s}{2(s-1)} \tag{5.2}$$

This is one of four canonical choices of diagonalization of the translation generators as may be seen by expanding the expression out more explicitly in dimensioned coordinates by using $\{x^a\} \simeq \{t/\lambda_t, q/\lambda_q\}$ where, for the moment, we do not assume natural units $c = b = \hbar = 1$ and so make the scales explicit.

$$\left(-t^2 + \frac{1}{c^2} q^2 - \frac{\hbar^2}{c^2} \frac{\partial^2}{\partial q^2} + \hbar^2 \frac{\partial^2}{\partial t^2} - \frac{\hbar}{cb}(2k + n - 1)\right)\psi_k(t, q) = 0$$

with solutions as above of the form

$$\psi_M(t, q) = \eta_k(t, q) = \eta_{(-k_0)}(t/\lambda_t)\, \eta_{k_1}(q^1/\lambda_q)\, \eta_{k_2}(q^2/\lambda_q) ... \eta_{k_n}(q^n/\lambda_q)$$

Where $k = k_0 +$ the function $\eta_{k_i}(s)\, i = 0, .. n$ is given in terms of the Hermite polynomials as

$$\eta_{k_i}(s) = \frac{1}{\sqrt{\pi\, 2^{k_i} k_i!}}\, e^{-\frac{1}{2} s^2} H_{k_i}(s), \, k_i \in \mathbb{Z}^+, \, s \in \mathbb{R}$$

Note that any of the diagonalizations may be chosen. We could repeat the above analysis with $\{X_a\} \simeq \{T, P_i\}$ and obtain the solution

$$\left(-t^2 + \tfrac{1}{b^2} p^2 - \tfrac{\hbar^2}{b^2} \tfrac{\partial^2}{\partial p^2} + \hbar^2 \tfrac{\partial^2}{\partial t^2} - \tfrac{\hbar}{cb}(2k+n-1)\right)\tilde{\psi}_k(t,p) = 0$$

As the $\eta_i(s)$ are their own Fourier transform, this result too is expressed in terms of the $\eta_i(s)$ functions

$$\psi_M(t, p) = \eta_k(t, q) = \eta_{(-k_0)}(t/\lambda_t)\, \eta_{k_1}(p^1/\lambda_p)\, \eta_{k_2}(p^2/\lambda_p) \dots \eta_{k_n}(p^n/\lambda_p)$$

In this theory, the usual concept of mass $\mu$ is not a Casimir invariant. Therefore, within a single irreducible representation of $C(1, n)$, there is a spectrum of particles with different masses. To compute the spectrum of the scalar particles, it is useful to consider the representation with $\{E, P_i\}$ diagonal (although of course the same computation can be done in any of the others). Then,

$$\left(-e^2 + b^2 p^2 - b^2 \hbar^2 \tfrac{\partial^2}{\partial p^2} + \hbar^2 \tfrac{\partial^2}{\partial e^2} - \hbar c b(2k+n-1)\right)\psi_k(e, p) = 0$$

This may be composed as

$$(\Box_\theta - j(j+n-2))\psi(\theta_1, \dots \theta_{n-1}) = 0,\ k \in \mathbb{Z} \geq 0,$$

$$\left(-\tfrac{\partial^2}{\partial \mu^2} - \tfrac{n}{\mu}\tfrac{\partial}{\partial \mu} + \tfrac{j(j+n-2)}{\mu^2} + \mu^2 - (2k+n)\right)\psi(\mu) = 0$$

where $\mu = \tfrac{1}{\lambda_e^2}(e^2 - c^2 p^2) > 0$ is mass as a dimensionless quantity in the natural scales, assumed to be greater than 0 and $\Box_\theta$ is the hyperbolic D'Alembertian on $\mathbb{H}_n^-$. This has the solution in terms of hyperbolic spherical functions $Y^j_{m_1..m_{n-1}}$ and Legendre polynomials $L^j_k$ [7] . Now, as we noted in the discussion for the corresponding Euclidean case, the mass operator is not an eigenvalue. This is quite different from the Poincaré theory where the Klein-Gordon equation for a spin 0 has the mass as a simple eigenvalue,

$$(\Box - \mu^2)\psi(x) = 0$$

Thus, in this theory, mass is only statistically given as an observable with probability distribution $\psi^\dagger(\mu)\,\mu\,\psi(\mu)$. Thus, mass is not confined to a mass hypershell given by the unitary irreducible representations of the Poincaré group but is a dynamical variable in this theory with transition probabilities that, in principle, are computable.

5.3.2 Representation of the fourth order Casimir invariant

The third equation $\alpha = 4$ for which $C_4 = W^2 = \eta^{a,b}\,\eta^{c,d}\,W_{a,c}\,W_{d,b}$ is

$$\begin{pmatrix}\left(a\!\left(x^a x^b - \tfrac{\partial^2}{\partial x^a \partial x^b}\right)\delta_{\tilde{M},M} - c_1 \Sigma_{a,b,\tilde{M},M}\right)\\ \left(a\!\left(x^b x^a - \tfrac{\partial^2}{\partial x^b \partial x^a}\right)\delta_{\tilde{M},M} - c_1 \Sigma_{b,a,\tilde{M},M}\right)\end{pmatrix}\psi_{\tilde{M}}(q) = c_4\,\psi_M(q)$$

This process may be repeated until the full set $\alpha = 1, 2, 4, 6, \ldots 2(n+1)$ is obtained for the $n+2$ independent Casimir invariants of $C(1, n)$.

The fourth order invariant may be recast in the form of the compact case (4.15) so that

$$\sum_{a,b\,M,\tilde{M}} (x^a - \tfrac{\partial}{\partial x^a})(x^b + \tfrac{\partial}{\partial x^b}) \psi_{\tilde{M}}(x) = \sum_{\alpha=1}^{2} \sum_{\kappa=0}^{2-\alpha+1} f_{\kappa,\alpha}^2(n+1, c_1, c_2, c_4) d_\alpha^{\;\kappa} \psi_M(x) \quad (5.3)$$

with the functions $f_{\kappa,\alpha}^2$ defined in (3.8).

The full solution of these and the higher order eigenvalue equations leads to the set of field equations for the canonical theory. We have previously shown that the quaplectic group contracts to the Poincaré group (plus the additional generators noted) in (3.11). A subsequent paper will analyze the specific form of these equations for the case $n = 3$ and the manner in which they contract in the limit $b \to \infty$ to the standard equations of the Poincaré theory.

# 6: Discussion

The above analysis demonstrates that unitary irreducible representations for the quaplectic group $C(1, n)$ can be constructed using the Mackey representation theory. Furthermore, the eigenvalue equations for the representations of the Casimir invariants define a set of field equations. In a coordinate basis, these field equations are second order differential equations and in the basis with position and time diagonal, act on a wave equation of the form $\psi_M(t, q)$. The second order Casimir equation is the familiar relativistic oscillator.

These higher order equations are rather complex and the analysis of them is barely sketched. Why should these equations be studied further?

The theory described is for a strongly interacting system that have relative forces of interaction are in the Planck scale. By this, we mean that the parameter $\gamma^i$ associated with the representation of the generator $\varrho'(N_i)$ is large or $f^i \to b$. As the action is only defined on the unitary representations of the group, a *basis* of which are the irreducible representations, the concept of force (and velocity) is only relative to these representations. How these manifest in the aggregated macroscopic classical limit of our everyday perception is a separate question.

A representation exists in which the values of the Casimir invariants $c_\alpha$ are all discrete. We say that it has a discrete spectrum. As particle properties, mass, lifetimes, intrinsic symmetries and so forth are generally observed in nature to be discrete. Why this representation would be singled out has not been determined. However, the existence of the representation is an encouraging step from the Poincaré group which only has a continuous mass spectra.

The $3 + 1$ Poincaré group is defined on the $\mathcal{L}^2$ Hilbert spaces over the 3 dimensional hypersurfaces $\mathbb{H}^3_{\pm,0}$ and each irreducible representation is labeled by a continuous mass eigenvalue of the representation of the second order Casimir operator. This is very concerning as mass transitions are observed in nature. The quaplectic group is the $\mathcal{L}^2$ Hilbert spaces over the full $\mathbb{R}^4$. These more general mass states are not eigenvalues of the representation of a Casimir invariant but rather are defined by the field equations that have a discrete spectra.

The theory hypothesizes that forces do not add with the usual Euclidean addition laws but through the action of the quaplectic group. The quaplectic group contains four distinct Poincaré subgroups. The Lorentz group of two of these are associated with velocity boosts, either on position-time or momentum-energy subspaces. The other two Lorentz groups define the generalized manner in which the forces add as that follows directly from (1.14) and (1.17) . The notions corresponding to classical *three* velocity and force are expressed in terms of hyperbolic tangents and are bounded.

There is a considerable body of literature associated with theories with a maximal acceleration, as for example described in the references of [45]. One of the goals is to address the infinities encountered in the quantum theories. One of the most notable of these is [44] as it seeks to provide a deep geometric foundation.

The theory described is a candidate to address the infinities encountered. As noted in the introduction, the author conjectures that the infinities may be associated with the limiting behavior $b \to \infty$ that causes the Hilbert space $\mathcal{L}^2(\mathbb{R}^4, \mathbb{R})$ to degenerate onto the hyperboloids $\mathcal{L}^2(\mathbb{H}^3_{\pm,0}, \mathbb{R})$. This requires further investigation.

The theory manifests Born's reciprocity principle. The theory is formulated over a nonabelian space $\mathbb{Q} = C(1, n)/\mathcal{SU}(1, n)$ and one can choose to diagonalize any pair of the representations of $\{T, E, Q_i, P_i\}$ on the edges of the quad

$$\begin{array}{ccc}
\varrho'(T)|\psi\rangle & \leftrightarrow & \varrho'(Q_i)|\psi\rangle \\
\updownarrow & & \updownarrow \\
\varrho'(P_i)|\psi\rangle & \leftrightarrow & \varrho'(E)|\psi\rangle
\end{array}$$

The field equations that are the representations of the Casimir invariants of the quaplectic group may be represented in a coordinate system corresponding to the diagonalization of $\{T, Q_i\}$, $\{E, P_i\}$, $\{T, P_i\}$ or $\{E, Q_i\}$. We are generally used to the first two diagonalizations in which the wave functions have the form $\psi_M(t, q)$ and $\psi_M(e, p)$ but in the quaplectic theory the representations $\psi_M(t, p)$ and $\psi_M(e, q)$ are equally relevant.

The theory that has been discussed is a global theory. That is, it is the counterpart of the special relativity theory as apposed to the general relativistic theory. It should be possible to create a more general theory by lifting the identification of $\mathbb{Q}$ to the tangent space of a general, nonabelian manifold with curvature and making the parameters local. Interestingly, Schuller [44] proved a *no go* theorem for Hermitian metrics of the form $-T^2 + \frac{1}{c^2} Q^2 + \frac{1}{b^2} P^2 - \frac{1}{b^2 c^2} E^2$ but that the extra $\frac{\hbar}{bc} I U$ term required for it to be a Casimir invariant of the quaplectic group (1.12) causes this *no go* theorem to not be applicable. Note in the limit $M_{a,b}$ transforms as an abelian (0, 2) tensor under the Poincaré group.

The dimensionality of the basic degrees of freedom and the explanation of all the three fundamental dimensional constants within the basic structure of the dynamical symmetry and not just $c$ in the context of the Poincaré dynamical symmetry and $\hbar$ in the context of the complete separate Heisenberg dynamical symmetry is remarkable.

The field equations that result from the quaplectic group are computationally complex. One would expect them to be so for a theory that has significant scope. But, the underlying principles are very simple and the theory is highly determined. Once the quaplectic group is adopted, there are very few arbitrary degrees of freedom. The Hilbert space of the unitary irreducible representations is completely determined. The field equations that are the eigenvalue equations of the representations of the Casimir operators are completely determined. Only a detailed analysis of the field equations that arise and there limiting behavior in the $b \to \infty$ where the quaplectic group contracts to the Poincare group will determine whether the quaplectic group is physically relevant.  This is the topic of a subsequent paper where the explicit form of the quaplectic field equations for $n = 3$ are obtained and the manner in which these equations and the  associated representation contract to the Poincaré field equations and associated representations are examined.

# 7: Notes and References

## 7.1 Notes

[1] E-mail: Stephen.Low@hp.com. v1.1 (8/14/04) has typographical errors corrected, notation changes, minor clarifications and material added on the $b \to \infty$ contraction limit added to v1.0 (4/30/04) in section 3.2.2. The notational change is the change of name of the group from *canonical* to quaplectic[2]. The title is left the same for consistencies - this paper will be condensed into shorter exposition suitable for journal publication.

[2] The term *canonical* has many existing meanings and feedback is that it is too general a term in this context and so this version changes the name of the group to *quaplectic*. *Quaplectic* is a word derived from the origins of the group in the *symplectic* group with the *quantum* (as in Heisenberg 'translation' subgroup) and *quad* (as in *quad* Poincaré subgroups and the nonabelian *quad*) connotations. The literal meaning of the word *qua* is similar to the preposition *as*, ' in the role or character of' and -*plectic* has origins in *pleat* 'to fold on itself'. So the literal meaning of *quaplectic* is 'in the character of folding on itself'. The author feels that this is a highly appropriate way of describing the relationship of the two relativities through the Born automorphisms.

[3] The physics convention is to represent the elements of the algebra with Hermitian operators (with real eigenvalues) rather than with anti-Hermitian with purely imaginary eigenvalues. As noted in the introduction, Hermitian and anti-Hermitian operators are in one to one correspondence simply by multiplication by $i$, $H = i \varrho'(X)$. Now, to avoid carrying around the $i$ explicitly, we adopt the convention of representing the algebra by Hermitian operators and incorporate the $i$ in the definition of $\varrho'$ and speak of Hermitian representations of the algebra. This means, that group an algebra elements are now related by the equation $\varrho(g) = e^{-i\varrho'(X)}$ for $g$ in the neighborhood of the identity. Furthermore, if the elements of the algebra satisfy the commutation relation, $[X, Y] = Z$, then the Hermitian representation satisfies $[\varrho'(X), \varrho'(Y)] = i\ \varrho'(Z)$.

[4] The number of Casimir invariants $N_c$ is the number is the multiplicity of zero $\lambda$ roots of the equation $\det[\![ c_{b,c}^a X^c - \lambda \delta_b^a ]\!] = 0$ where $c_{b,c}^a$ are the structure constants of the algebra.

[5] The term 'matrix group that are *algebraic'* to describe closed subgroups of the general linear group through polynomial constraints originates with Chevalley. See also [17] pg 85 and [49].

[6] This concept of the dynamical group is motivated by the work of Barut and Bohm [6]. We take the active dynamical view of the Poincaré transformations rather than the passive symmetry view. The group yields the basic dynamical equations of the particle field equations as noted.

[7] The strict definition of the Poincare group is the cover of the group $\mathcal{E}(1, n)$. For $n = 3$, $\mathcal{E}(1, 3) \simeq \mathcal{SL}(4, \mathbb{C})$.

[8] We use the notation $\mathcal{T}(1, n)$ and $\mathcal{H}(1, n)$ even though $\mathcal{T}(1, n) \simeq \mathcal{T}(1 + n)$ and $\mathcal{H}(1, n) \simeq \mathcal{H}(1 + n)$ as convenient way of keeping track of the signature of a metric that is in the case under consideration. Note that the translation and Heisenberg group have no concept of an orthogonal metric which is only introduced by the orthogonal group

[9] A factor of $i$ appears in the definition of the Heisenberg algebra as follows. The conventions of Dirac are that the commutators of the Hermitian representation of the position and momentum generators are $[\varrho'(P_i), \varrho'(Q_i)] = i\,\delta_{i,j}\,\varrho'(I)$. Then noting comments in note [3] it follows that the real form of the abstract algebra that $[P_i, Q_i] = \delta_{i,j}\,I$ and consequently the complex formula has $[A_i^+, A_j^-] = i\,\delta_{i,j}\,I$

[10] See Folland [15] pg 20. $\mathcal{Sp}(2n) \otimes_s \mathcal{H}(n)$ is a subgroup of the group of automorphisms. In addition there are the dilations and the discrete reflections of the center that Folland describes.

[11] The algebra contractions are described in Gilmore [17], chapter 10 and can be formalized to the degree required.

[12] Turning the equation around so that $b$ is more basic, $G = \alpha_G \frac{c^4}{b}$ where a dimensionless constant $\alpha_G$ is inserted. If this constant is not unity, then a fully complete theory would need to compute it just as it should compute other dimensionless coupling constants such as the fine structure constant that defines the charge coupling and so forth.

[13] Note that we use the same symbol $\varrho$ for the unitary irreducible representations determined from the Mackey theory as in the Poincaré case without an explicit label to distinguish them. Carrying a label such as $\varrho(\mathcal{C}(1, n))$ becomes somewhat baroque. The representation should be clear from the context.

## 7.2 References

[1] G. Amelino-Camelia, *Planck-scale structure of spacetime and some implications for astrophysics and cosmology*, Preprint (2003). arXiv: gr-qc/0312014

[2] G.C. Baird and L.C. Biedenharn, *On the Representations of the Semisimple Lie Groups. II*, J. Math. Phys., **4**, 1449-1466 (1963).

[3] A. Baker, *Matrix Groups; An Introduction to Lie Group Theory,* (Springer, London, 2002).

[4] S. Barbari and C.S. Kalman, $SU(1, 3)$ *as a Dynamical Group for Hadron Scattering States*, L. al N. Cim. **27**, 513-518 (1980).

[5] V. Bargmann, *On a Hilbert Space of Analytic Functions and an Associated Integral Transform*, Comm. P. Apply. Math., **14**, 187-214 (1961).

[6] A, Bohm, *Relativistic and Nonrelativistic Dynamical Groups,* Found. Phys. , **23**, 751-767 (1993).

[7] M. Born, *A suggestion for unifying quantum theory and relativity*, Proc. Roy. Soc. London, A165, 291-302 (1938)

[8] M. Born, *Reciprocity Theory of Elementary Particles*, Rev. Mod. Phys., **21**, 463-473 (1949).

[9] E.R. Cainiello, *Is there a Maximal Acceleration*, Il. Nuovo Cim., **32**, 65-70 (1981).

[10] M. Chaichian, A.P. Demichev and N.F. Nelipa, *The Casimir operators of inhomogeneous groups*, Commun. Math. Phys., **90**, 353-372 (1983).

[11] P. A. Cook, *Relativistic Harmonic Oscillators with Intrinsic Spin Stuctures,* Let. Al N. Cimento, **1**, 419-426 (1971), *Is there a Maximal Acceleration*, Il. Nuovo Cim., **32**, 65-70 (1981).

[12] P.A.M Dirac, *Quantum Mechanics*, (Oxford University Press, Oxford, 1958).

[13] J.P. Elliot and P.G. Dawber, *Symmetry in Physics,* Vol I & II, (Oxford University Press, Oxford, 1979).

[14] A. Feoli, *Maximal acceleration or maximal accelerations*, Int. J. Mod. Phys., D12, 271- 280 (2003). arXiv: gr-qc/0210038